\documentclass[prx,aps,twocolumn,longbibliography,superscriptaddress]{revtex4-2}

\usepackage[pdftex]{graphicx}
\usepackage{amsbsy,amssymb,amsmath,bm,mathtools}

\usepackage{bm}% bold math
\usepackage{color}
 
\usepackage{url}
\usepackage[section]{placeins}
\usepackage[colorlinks=true,linkcolor=blue,citecolor=blue]{hyperref}

\usepackage{comment}

\begin{document} 
\let\vec\mathbf
%\preprint{Preprint}

\title{Machine learning force-field models for metallic spin glass}

\author{Menglin Shi}
\affiliation{Department of Physics, University of Virginia, Charlottesville, VA 22904, USA}

\author{Sheng Zhang}
%\thanks{ORCID:}
\affiliation{Department of Physics, University of Virginia, Charlottesville, VA 22904, USA}

\author{Gia-Wei Chern}
\affiliation{Department of Physics, University of Virginia, Charlottesville, VA 22904, USA}
%\thanks{gchern@virginia.edu }

\date{\today}

\begin{abstract}
Metallic spin glass systems, such as dilute magnetic alloys, are characterized by randomly distributed local moments coupled to each other through a long-range electron-mediated effective interaction. We present a scalable machine learning (ML) framework for dynamical simulations of metallic spin glasses. A Behler-Parrinello type neural-network model, based on the principle of locality, is developed to accurately and efficiently predict electron-induced local magnetic fields that drive the spin dynamics. A crucial component of the ML model is a proper symmetry-invariant representation of local magnetic environment which is direct input to the neural net. We develop such a magnetic descriptor by incorporating the spin degrees of freedom into the atom-centered symmetry function methods which are widely used in ML force-field models for quantum molecular dynamics.  We apply our approach to study the relaxation dynamics of an amorphous generalization of the s-d model. Our work highlights the promising potential of ML models for large-scale dynamical modeling of itinerant magnets with quenched disorder. 
\end{abstract}

% As the sign of the effective interaction randomly oscillates between ferro- and antiferromagnetic couplings, interaction energy between spin pairs cannot be simultaneously minimized. The resultant magnetic frustration thus precludes long-range order, giving rise to disordered, yet frozen, spins at low temperatures.  

\maketitle

\section{introduction}

\label{sec:intro}

Spin glasses are disordered magnetic materials in which interactions between localized magnetic moments are frustrated due to quenched disorder~\cite{mezard87,binder86,mydosh95,mydosh15}. As a result of the frustrated interactions, no conventional long-range magnetic order, such as the ferromagnetic or N\'eel states, can be established. Yet spin glass systems exhibit collective freezing transitions below a characteristic temperature $T_f$, indicating new magnetic states of matter at low temperatures.  The term ``glass'' comes from the analogy between the disordered spins in a low-temperature spin-glass phase and  the atomic positional disorder of a structural glass~\cite{dyre06,charbonneau17}. It is worth noting that, whereas a frozen disordered atomic configuration arises spontaneously from the exponentially slow relaxation of a supercooled liquid, quenched randomness plays a central role in spin glass systems. 

Several theoretical models have been introduced to understand the nature of spin-glass phases and transitions. A canonical example is the Edwards-Anderson model~\cite{edwards75} which describes random exchange interactions between nearest-neighbor Ising spins $\sigma_i$ on a lattice: $\mathcal{H} = -\sum_{\langle ij \rangle} J_{ij} \sigma_i \sigma_j$. The coupling coefficients are random variables with equal probability of being positive (ferromagnetic) or negative (antiferromagnetic). Mean-field theories of spin glasses, such as the Sherrington-Kirkpatrick model~\cite{sherrington75}, further shed light on the nature of spin-glass ``order", which later led to the idea of replica symmetry breaking and related order parameters for spin-glass phases~\cite{thouless77,parisi79,parisi83,mezard84}. The physical insight and mathematical techniques developed in the study of spin glass have found applications in disciplines as diverse as computer science, biology, and economics~\cite{nishimori01,stein13,charbonneau23}. 

% Frustration occurs when the product of $J_{ij}$ around a closed loop of nearest-neighbor spins is negative, since interactions between spin-pairs along the loop cannot be simultaneously satisfied~\cite{toulouse77}. 

% Spin-glass systems have become a paradigm for the understanding of glassy materials. 

Experimentally, the prototype spin-glass materials are dilute magnetic alloys, where a small amount of magnetic impurity, such as Fe or Mn, are randomly substituted into the lattice of a nonmagnetic metallic host (e.g. Ag, Cu, Pt)~\cite{binder86,mydosh95,mydosh15,cannella72,lamelas95}. They are prepared by rapidly cooling the liquid alloy, thus fixing the strongly interacting particles at random positions within the resulting solid. As these magnetic impurities are typically several lattice constants away from each other, their effective interactions are mediated by conducting electrons from the metallic host. Due to the itinerant nature of conducting electrons, the resultant effective spin-spin interactions are usually long-ranged, as exemplified the by well-known Ruderman-Kittel-Kazuya-Yosida (RKKY) interactions~\cite{ruderman54,kasuya56,yosida57}. The RKKY interaction also oscillates between ferro- and antiferromagnetic couplings with increasing distances, offering a frustration mechanism for stabilization of spin-glass phases. 

The RKKY type interactions, however, are obtained based on the assumption of a spherical Fermi surface in the limit of weak electron-spin coupling. In general, the electron-mediated interactions in metallic spin glasses depend on the electronic structure of the underlying metallic system. A full dynamical simulation of such disordered itinerant electron magnets, however, is  computationally a very challenging task. As the driving forces come from electrons, integration of the Landau-Lifshitz-Gilbert (LLG) equation for spin dynamics requires solving an electron Hamiltonian at every time-step. Such repeated electronic structure calculations could be prohibitively time-consuming for dynamical simulations of large systems. The computational complexity is similar to the quantum molecular dynamics (QMD) simulations, where atomic forces are computed from electronic structure solutions~\cite{marx09}. 

In this paper, we present a scalable machine learning (ML) framework for large-scale dynamical modeling of metallic spin glass and disordered itinerant electron magnets. Central to our approach is a deep-learning neural network, trained by exact solutions from small systems, that can accurately and efficiently predict the effective local magnetic fields acting on spins. This approach is similar in spirit to ML-based force field models which have revolutionized the field of {\em ab initio} MD simulations~\cite{behler07,bartok10,li15,shapeev16,botu17,smith17,zhang18,behler16,deringer19,mcgibbon17,suwa19,chmiela17,chmiela18,sauceda20}. By accurately emulating the calculation of Kohn-Sham equations, ML interatomic potentials offer the efficiency of classical force-field models with the desired quantum accuracy. Similar ML frameworks have recently been developed to enable large-scale dynamical simulations in several condensed-matter lattice systems~\cite{zhang20,zhang21,zhang22,zhang22b,zhang23,cheng23,cheng23b}. 

Physically, the linear scalability of ML force-field models is based on the principle of locality, or the nearsightedness, of electronic matter~\cite{kohn96,prodan05}. A practical approach to incorporate the locality principle into ML models for quantum MD was proposed in the pioneering work of Behler and Parrinello~\cite{behler07}. The central idea of Behler-Parrinello (BP) type ML structures involves introducing a local energy $\epsilon_i$~\cite{behler07,bartok10} which, in our case, is associated with individual spins $\mathbf S_i$ in a disordered magnet. The local effective field is obtained from the derivative of the total energy. Importantly, invoking the locality principle, the local energy is assumed to depend on the immediate neighborhood of $\mathbf S_i$ and a deep-learning neural net is trained to capture this complex dependence.  

A crucial component of the scalable ML model is the appropriate representation of the local environment which preserves the symmetry of the original quantum systems. Indeed, atomic descriptors play a central role in the field of ML-based quantum MD methods~\cite{behler07,bartok10,li15,behler11,ghiringhelli15,bartok13,drautz19,himanen20,huo22}.  In our case, a proper representation of spin configurations in a local neighborhood needs to be invariant with respect to real-space rotation and translation symmetries, as well as the SU(2) rotation symmetry in the spin-space. To this end, we develop a magnetic descriptor which is a generalization of a widely used atomic descriptor called atom-centered symmetry function (ACSF) for ML-MD simulations~\cite{behler07,behler11}. 

We apply our ML force-field model to study the relaxation dynamics of a metallic spin-glass model with Heisenberg spins. We note that most works on spin-glass phases are based on lattice models of either Ising or Heisenberg spins with quenched random nearest-neighbor exchange interaction $J_{ij}$~\cite{binder86}. There are a few remarkable works on the {\em ab initio} modeling and dynamical simulations of the prototypical MnCu metallic spin-glass alloys~\cite{skubic09,hellsvik10,peil08,ling94}. In these approaches, however, a classical Heisenberg model is used to describe the magnetic alloys with an RKKY-like exchange interaction determined from first-principles density functional theory calculations~\cite{ruban04}. To demonstrate the ML model for electron-driven spin dynamics, here we consider an amorphous generalization of the well-studied s-d model as a model metallic spin glass. Exact diagonalization (ED) combined with LLG dynamics are used to simulate this random s-d model and to generate datasets for training of the ML models.

The outline of the paper is as follows. In Sec.~\ref{sec:dynamics} we first present a generic model for metallic spin glass based on a local s-d type electron-spin coupling. Magnetization dynamics of itinerant electron magnets in the adiabatic approximation is then discussed. in Sec.~\ref{sec:ML-model}, we outline the general ML framework for efficient computation of the electron-induced local magnetic fields, similar to atomic forces in quantum MD simulations. The ML approach is applied to a specific random s-d model and the benchmarks with ED calculations are presented in Sec.~\ref{sec:results}. The ML-based LLG simulations are carried out to study of relaxation dynamics of the amorphous s-d model.  Finally, Sec.~\ref{sec:conclusion} concludes the article with a summary and outlook.

\section{Adiabatic dynamics of itinerant electron magnets}

\label{sec:dynamics}

As discussed in several previous works, magnetic moments in dilute magnetic alloys such as CuMn can be well described by Heisenberg spins. We consider a metallic spin system with randomly distributed Heisenberg spins, based on the s-d type electron-spin coupling~\cite{anderson61}:
\begin{eqnarray}
	\label{eq:H1}
	\hat{\mathcal{H}} = \hat{\mathcal{H}}_e\left(\hat{c}, \hat{c}^\dagger \right) - J \sum_i \sum_{\alpha,\beta = \uparrow,\downarrow} \mathbf S_i \cdot   \left(\hat{c}^\dagger_{i, \alpha} \bm\sigma^{\,}_{\alpha\beta} \hat{c}^{\,}_{i, \beta} \right).
\end{eqnarray}
Here $\mathbf S_i$ represents a local classical spin at a random position $\mathbf r_i$, and $c^\dagger_{i, \alpha}$ ($c^{\,}_{i, \alpha}$) denotes the creation (annihilation) operator of an electron with spin $\alpha = \uparrow, \downarrow$ localized at $\mathbf r_i$. The first term $\hat{\mathcal{H}}_e$ is the Hamiltonian of the electron subsystem, which corresponds to the metallic host in dilute magnetic alloys. The electron operators in the parentheses of the second term correspond to the spin operator of the electrons: $\hat{\mathbf s}_i = \frac{\hbar}{2} \left(\hat{c}^\dagger_{i, \alpha} \bm\sigma_{\alpha\beta} \hat{c}^{\,}_{i, \beta} \right)$. Physically, the coefficient $J$ describes a Hund's coupling between electron spin and local moment. 

As discussed in Sec.~\ref{sec:intro}, in the limit of small electron-spin coupling, one can integrate out the electrons to obtain an effective interaction between local spins. In particular, assuming an electronic system described by a Fermi liquid $\hat{\mathcal{H}}_e = \sum_{|\mathbf k| < k_F}\sum_\alpha \varepsilon_{\mathbf k} \hat{c}^\dagger_{\mathbf k, \alpha} c^{\,}_{\mathbf k, \alpha}$, with a parabolic dispersion $\varepsilon_{\mathbf k} = \hbar^2 |\mathbf k|^2 / 2m_e$, one obtains the well-known RKKY interaction~\cite{ruderman54,kasuya56,yosida57}
\begin{eqnarray}
	\label{eq:RKKY}
	E = \sum_{ij} J_{\rm RKKY}(2 k_F |\mathbf r_i - \mathbf r_j|)  \mathbf S_i \cdot \mathbf S_j,
\end{eqnarray}
where $J_{\rm RKKY}(x) = 9\pi (J^2/\varepsilon_F)  (x \cos x - \sin x)/x^4$, $k_F$ and $\varepsilon_F$ are the Fermi wavevector and Fermi energy, respectively.   For large coupling $J \gtrsim \varepsilon_F$ and more complex electronic models, the effective interaction in general needs to be computed numerically. For example, such effective interactions have been computed using DFT for realistic CuMn alloys~\cite{skubic09,hellsvik10,peil08}.

The magnetization dynamics for Heisenberg magnets is governed by the Landau-Lifshitz-Gilbert (LLG) equation
\begin{eqnarray}
	\label{eq:LLG}
	\frac{d\mathbf S_i}{dt} = \gamma \mathbf S_i \times ( \mathbf H_i + \bm\eta_i )  
	- \alpha \mathbf S_i \times (\mathbf S_i \times \mathbf H_i),
\end{eqnarray}
where $\gamma$ is the gyromagnetic ratio, $\alpha$ is the damping constant, $\mathbf H_i$ is the local effective field that drives the spin dynamics, and $\bm\eta_i$ represents a stochastic magnetic field due to thermal fluctuations. The random fields at different sites and times are uncorrelated. The three independent components of $\bm\eta_i$ are modeled by a Gaussian random variable of zero mean and a variance proportional to $\alpha k_B T$, where $T$ is temperature. For effective spin Hamiltonian, such as the RKKY interaction in Eq.~(\ref{eq:RKKY}), the effective local field is given by
\[
	\mathbf H_i = -\frac{\partial E}{\partial \mathbf S_i} = - \sum_j J_{\rm RKKY}(2 k_F |\mathbf r_i - \mathbf r_j|  ) \mathbf S_j
\]
Although this interaction is long-ranged, the $1/r^3$ decay is sufficiently fast that a cutoff radius is introduced in practical calculations. The effective field in general can be efficiently computed for such RKKY type interactions.

It is worth noting that spin dynamics based on such effective interactions is a special case of the adiabatic approximation, which is similar to the Born-Oppenheimer approximation for quantum MD simulations~\cite{marx09}. Within the adiabatic approximation, electrons are assumed to quickly relax to quasi-equilibrium with respect to the instantaneous configuration of classical spins. In this limit, the effective energy is given by $E = \langle \hat{\mathcal{H}} \rangle = {\rm Tr}( \hat{\rho}_e \hat{\mathcal{H}})$, where $\hat{\rho}_e$ is the electron density operator:
\begin{eqnarray} 
	\label{eq:rho_e}
	\hat{\rho}_e = \exp(-\hat{\mathcal{H}}/k_B T)/\mathcal{Z},
\end{eqnarray} 
and $\mathcal{Z} = {\rm Tr} \exp(-\hat{\mathcal{H}}/k_B T)$ is the partition function. The local magnetic field for Hamiltonian~(\ref{eq:H1}) in this adiabatic limit is given by
\begin{eqnarray}
	\mathbf H_i = - \frac{ \partial \langle \hat{\mathcal{H}} \rangle}{\partial \mathbf S_i}  
	= J \sum_{\alpha\beta}\bm \sigma^{\,}_{\alpha\beta} \langle \hat{c}^\dagger_{i, \alpha}  \hat{c}^{\,}_{i, \beta} \rangle.
\end{eqnarray}
Here we have used the Hellmann-Feynman theorem $\partial \langle \hat{\mathcal{H}} \rangle / \partial \mathbf S_i = \langle \partial \hat{\mathcal{H}} / \partial \mathbf S_i \rangle$ to relate the local field to the electron correlation function $C_{i\alpha, j \beta} \equiv \langle \hat{c}^\dagger_{j, \beta}  \hat{c}^{\,}_{i, \alpha} \rangle$. The calculation of the correlation function, however, requires solution of the equilibrium electron density matrix $\hat{\rho}_e$, which amounts to solving a disordered electron Hamiltonian for a given spin configuration. In the absence of electron-electron interactions, the standard method for solving the electronic structure is based on exact diagonalization (ED). However, since the electronic forces have to be computed at every time-step of the LLG dynamics simulation, the $\mathcal{O}(N^3)$ time complexity of ED can be overwhelmingly time-consuming for large systems.

\section{Machine learning force-field model for disordered spins}

\label{sec:ML-model}

In this section we present a scalable ML framework to essentially derive an effective spin Hamiltonian $E(\mathbf S_i)$ for metallic spin glass models in Eq.~(\ref{eq:H1}). It should be noted that this effective energy $E$, or effective classical spin Hamiltonian, is technically obtained by freezing the spin configuration and integrating out the electrons using the electron density operator in Eq.~(\ref{eq:rho_e}). It can be viewed as a more complicated version of the RKKY interaction. The ML methods offer a systematic approach to obtain an accurate and efficient parametrization of this effective spin Hamiltonian. 

\subsection{Behler-Parrinello ML framework}

Fundamentally, as discussed in Sec.~\ref{sec:intro}, linear-scaling electronic structure methods are possible mainly because of the locality nature of many-electron systems~\cite{kohn96,prodan05}. Modern ML techniques provide an explicit and efficient approach to incorporate the locality principle into the implementation of $\mathcal{O}(N)$ methods. In particular, the Behler-Parrinello type schemes provide a practical method to incorporate locality and symmetry to the ML models. Indeed, most ML force-field models for quantum MD simulations are based on BP approaches. Here we generalize the BP scheme to implement an ML model for the efficient prediction of system energy $E$ and the local effective fields $\mathbf H_i$ for spin dynamics. 

\begin{figure*}[t]
\centering
\includegraphics[width=1.99\columnwidth]{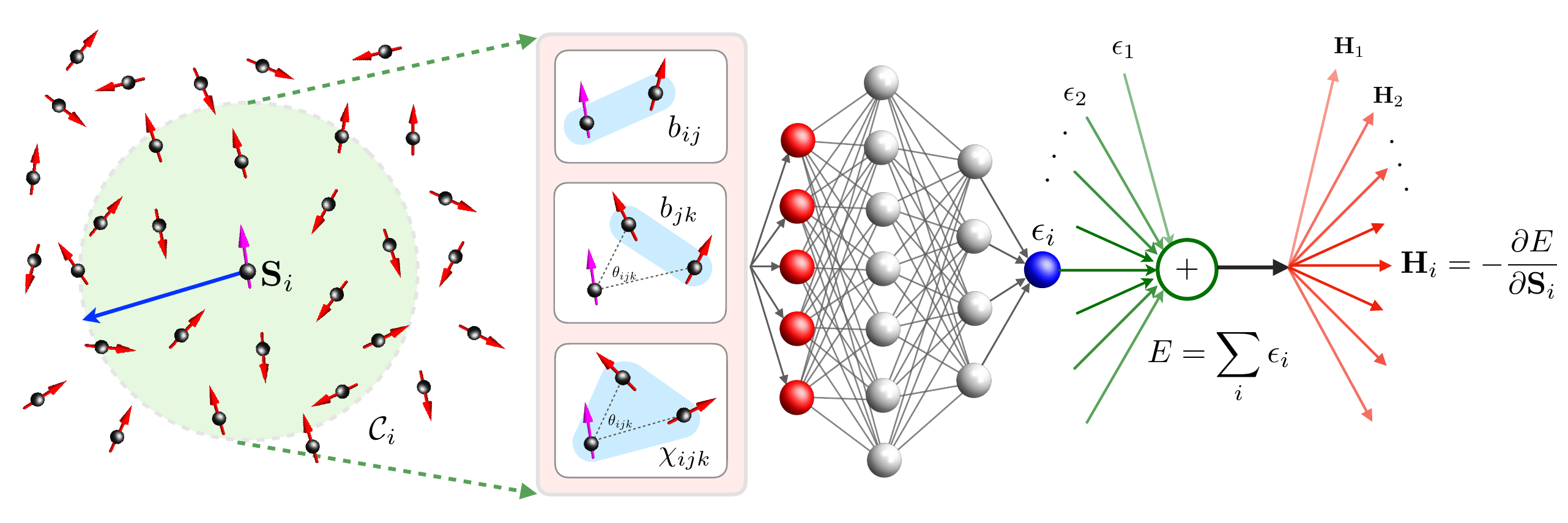}
\caption{Schematic diagram of ML force-field model for metallic spin glass. The input of the ML model is the magnetic configuration $\mathcal{C}_i$ centered at the $i$-th spin $\mathbf S_i$, while the output is the local energy $\epsilon_i$ associated with the center spin. The atomic and spin configuration within the neighborhood $\mathcal{C}_i$ is processed using a magnetic version of the ACSF method. The resultant feature variables $\{ G_k \}$ are input to a feed-forward fully connected neural network with a single output that is the local energy $\epsilon_i$. The total energy is obtained by applying the ML energy model to all spins in the system. The local effective field is computed using automatic differentiation. }
    \label{fig:ml-model}
\end{figure*}

A schematic diagram of our ML force-field model is outlined in FIG.~\ref{fig:ml-model}. First, as in the original BP approach, we partition the effective spin energy into local contributions 
\begin{eqnarray}
	\label{eq:E_ML}
	E = \sum_i \epsilon_i = \sum_i \varepsilon(\mathcal{C}_i), 
\end{eqnarray}
where $\epsilon_i$ is associated with the $i$-th local spin of the disordered magnet. In the second step of the above equation, we have further assumed that the local energy only depends on the immediate neighborhood, denoted by $\mathcal{C}_i$, of the $i$-th spin through a universal function $\varepsilon(\cdot)$. The validity of this step naturally relies on the locality principle mentioned above. Practically, the neighborhood $\mathcal{C}_i$ is defined as configuration of spins within a sphere of cutoff radius~$R_c$ centered at $\mathbf S_i$, i.e. $\mathcal{C}_i = \left\{ \mathbf S_j \, \big| \, |\mathbf r_j - \mathbf r_i| < R_c \right\}$; see FIG.~\ref{fig:ml-model}. Importantly, the universal function $\varepsilon(\cdot)$ that relates local energy $\epsilon_i$ to the neighborhood is to be approximated by a deep-learning neural network model. As shown in FIG.~\ref{fig:ml-model}, the local magnetic environment is processed to produce a set of symmetry-invariant feature variables, denoted as $\{G_m\}$, which are then fed into the neural network. 

Once the total energy $E$ is obtained by applying the same descriptor and neural network model to every spin in the system, the local field acting on spin $\mathbf S_i$ can then be efficiently computed through automatic differentiation:
\begin{eqnarray}
	\label{eq:H_ML}
	\mathbf H_i = -\frac{\partial E}{\partial \mathbf S_i} = - \sum_j\!^{'} \frac{\partial \epsilon_j }{\partial \mathbf S_i}.
\end{eqnarray}
Here the prime in the summation indicates that only local energies $\epsilon_j$ within the neighborhood of the $i$-th spin are considered, again, assuming the locality of the effective field. 

Another important insight in the original work of Behler and Parrinello~\cite{behler07} is the introduction of descriptors, or feature variables, that provide an appropriate representation of the neighborhood $\mathcal{C}_i$.  It is worth noting that, despite the universal approximation power of neural networks, symmetry properties of a function can only be learnt statistically, but not exactly. The goal of a descriptor is to incorporate the symmetry of the original quantum Hamiltonian into the classical effective energy model. Since the output of the ML model in the BP framework is a local energy, which, as a scalar, is invariant under symmetry transformations of the original system, the feature variables $\{G_m\}$ need to be also invariant with respect to the same symmetry group.

In the case of ML-based quantum MD methods, extensive studies have been devoted to the development of atomic descriptors~\cite{behler07,bartok10,li15,behler11,ghiringhelli15,bartok13,drautz19,himanen20,huo22}.
A proper representation of the atomic neighborhood should be invariant under rotational and permutational symmetries, while retaining the faithfulness of the Cartesian representation. A widely used atomic descriptor, which is physically intuitive, is the atom-centered symmetry function (ACSF) representation introduced in the original work of Behler and Parrinello~\cite{behler07}. The ACSFs are built from the relative distances and relative angles  among atoms in the neighborhood, which are manifestly invariant under rotations~\cite{behler07,behler11}. 
A more systematic approach to build invariant feature variables is based on the so-called bispectrum coefficients, which are special triple-products of irreducible representations (IRs) of the symmetry group~\cite{bartok10,bartok13,kondor07}. 

The group-theoretical method has also been employed to develop a general theory of descriptors for electronic lattice models in condensed-matter systems~\cite{zhang22}. Compared with the MD systems, the SO(3) rotational symmetry of free-space is reduced to discrete point-group symmetries in lattice models. On the other hand, the dynamical degrees of freedom in lattice models, such as local magnetic moments or order-parameters, are characterized by additional internal symmetry group. A proper descriptor for lattice systems thus needs to be invariant with respect to both the internal symmetry group and the lattice point group. 

In particular, following the general theory, a magnetic descriptor is developed to incorporate the lattice point-group and SU(2) spin-rotation symmetries into the ML force-field models for lattice models of itinerant electron magnets, such as the s-d or Kondo-lattice models~\cite{zhang20,zhang21,zhang23}. A two-step approach is used to construct the feature variables. First, to preserve the rotation symmetry in spin-space, bond $b_{jk} = \mathbf S_j \cdot \mathbf S_k$ and scalar chirality $\chi_{ijk} = \mathbf S_i \cdot \mathbf S_j \times \mathbf S_k$ variables, which are inner product and scalar triple product of spins within a neighborhood, are used as building blocks for the magnetic descriptor.  A group-theoretical approach based on reference IR~\cite{zhang22}, which is a modified bispectrum method, is used to derive feature variables that are also invariant under symmetry operations of the lattice point group. 

We note in passing that in order to model atomic systems with finite magnetic moments, atomic descriptors have been generalized in several recent works to include spin degrees of freedom~\cite{drautz20,eckhoff21,brannvall22,domina22,novikov22,chapman22}. For such magnetic materials, magnetic contributions are crucial for modeling the mechanic phase stability, vibrational properties, and defect dynamics.   However, since the goal of these descriptors is to assist ML-based quantum molecular dynamics, the magnetic structure of atoms are assumed to be fixed throughout the MD simulations. Moreover, often collinear Ising-type spins are considered. Here, on the other hand, we present a magnetic descriptor to be combined with an ML model for spin dynamics, whereas atomic configurations are fixed.  

\subsection{Magnetic descriptor for disordered spins}

\label{sec:descriptor}

The symmetry group of the spin glass model described in Eq.~(\ref{eq:H1}) includes the real-space translation and rotation symmetries of atoms (which carry a local spin $\mathbf S_i$) and the global SO(3)/SU(2) rotation symmetry of classical/electron spins. The translation symmetry is readily accounted for in the BP framework, as exactly the same descriptor and neural network are applied to every spin in the system. To account for the rotation symmetry of atoms in real space, we first review the idea of symmetry functions which are built on two fundamental scalars: the relative distance between two atoms $R_{ij} = |\mathbf r_j - \mathbf r_i|$ and the relative angle between three atoms $\cos\theta_{ijk} = (\mathbf r_j - \mathbf r_i) \cdot (\mathbf r_k - \mathbf r_i) / R_{ij} R_{ik}$.  As scalars, these two types of variables are manifestly invariant under uniform rotations of atoms in the neighborhood.  A two-body symmetry function centered at $\mathbf r_i$ is defined as
\begin{eqnarray}
	\label{eq:G2-original}
	G_2(\Lambda) = \sum_{j \neq i} F_2(R_{ij}; \, \Lambda),
\end{eqnarray}
where $\Lambda = \{p_1, p_2, \cdots\}$ denote a set of parameters characterizing $G_2$, and $F_2(R; \Lambda)$ is a user-defined function, parameterized by $\Lambda$ to extract atomic structures at certain distances from the center atom. For example, the following function is proposed in the original work~\cite{behler07} to sample atoms at a distance $d \pm w$ from the center
\begin{eqnarray}
	\label{eq:F2-envelop}
	F_2(R; d, w) = e^{- ( R - d)^2/w^2 }\, f_c(R ).
\end{eqnarray}
Here $f_c(r) = \frac{1}{2} \bigl[ \cos(\frac{\pi r}{R_c}) + 1 \bigr]$ for $R \le R_c$ and zero otherwise is a soft cutoff function. The two parameters $d$ and $w$ specify the center and width, respectively, of the Gaussian function. The 3-body symmetry functions are defined as
\begin{eqnarray}
	 G_3(\Lambda) &=& \sum_{j, k \neq i} F_3(R_{ij}, R_{ik}, R_{jk}, \theta_{ijk}; \, \Lambda),
\end{eqnarray}
An example of the three-body envelop function characterized by three parameters is~\cite{behler07,behler16} 
\begin{eqnarray}
	\label{eq:F3-envelop}
	& & F_3(R_1, R_2, R_3, \theta; \zeta, \lambda, d, w, d', w') =  2^{1- \zeta}  (1 + \lambda \cos\theta )^{\zeta}  \nonumber \\
	& & \qquad \times F_2(R_1; d, w) F_2(R_2; d, w) F_2(R_3; d', w') . 
\end{eqnarray}
These functions are designed to sample relative orientations between atomic pairs, where the angular resolution is controlled by the parameter $\zeta$. The three $F_2$ functions are introduced to constrain the distances of the three pairs of spins. 

To incorporate the spin degrees of freedom, we first note that assuming the electronic part $\mathcal{H}_e$ in Eq.~(\ref{eq:H1}) is magnetically isotropic, the metallic spin system is invariant under a global rotation of local magnetic moments $\mathbf S_i \to \mathcal{R}\cdot \mathbf S_i$, and a simultaneous unitary transformation of the electron spinor $\hat{c}_{i\alpha} \to \hat{U}_{\alpha\beta} \hat{c}_{i\beta}$, where $\mathcal{R}$ is an orthogonal $3\times 3$ matrix and $\hat{U} = \hat{U}(\mathcal{R})$ is the corresponding $2\times 2$ unitary rotation operator. The ML force-field model, which is essentially an effective classical spin model by integrating out electrons, should preserve the SO(3) spin-rotation symmetry. Similar to magnetic descriptors for lattice models, we use the bond and scalar spin chirality variables $b_{ij} = \mathbf S_i \cdot \mathbf S_j$ and $\chi_{ijk} = \mathbf S_i \cdot \mathbf S_j \times \mathbf S_k$ as building blocks for characterization of the magnetic environment~\cite{zhang20,zhang21,zhang23}. These variables are also manifestly invariant under global rotations of all spins. 

Based on these building blocks, invariants with respect to rotations in both real space and spin space are obtained by ``attaching" these variables into the symmetry functions. A schematic diagram showing the construction of the three types of magnetic symmetry functions is shown in FIG.~\ref{fig:ml-model}.  First, we define a magnetic two-body symmetry function
\begin{eqnarray}
	\label{eq:G2-spin}
	G^m_{2}( \Lambda ) = \sum_{j \neq i} F_{2}( R_{ij}; \, \Lambda  ) \,  (\mathbf S_i \cdot \mathbf S_j).
\end{eqnarray}
Here the superscript $m$ is used to indicate magnetic version of the ACSF.
With the $F_2$ function defined in Eq.~(\ref{eq:F2-envelop}), this magnetic symmetry function accounts for spin-spin correlations between the center $\mathbf S_i$ and neighboring spins $\mathbf S_j$ at a distance $d \pm w$ from the center. Next, a symmetry function which involves three atoms is defined:
\begin{eqnarray}
	\label{eq:G3a-spin}
	G^m_{3}( \Lambda ) = \sum_{jk \neq i} F_{3}\bigl(R_{ij}, R_{ik}, R_{jk}, \theta_{ijk}; \, \Lambda) \, (\mathbf S_j \cdot \mathbf S_k), \qquad
\end{eqnarray}
This symmetry function, however, only involves two spins. Collectively, they describe the two-spin correlations with certain relative angles and distances in the neighborhood of the $i$-th spin.

The total energy Eq.~(\ref{eq:E_ML}) computed from the ML model is now a function explicit of these magnetic ACSFs, i.e. $E = E(\{ G_2^m(\Lambda), G_3^m(\Lambda) \})$. This can also be viewed as an effective classical spin Hamiltonian, which can be formally expanded as
\begin{eqnarray}
	\label{eq:E_expansion}
	E = \sum_{ij} J_{ij} \, \mathbf S_i \cdot \mathbf S_j + \sum_{ijkl} K_{ijkl} (\mathbf S_i \cdot \mathbf S_j) (\mathbf S_k \cdot \mathbf S_l) + \cdots \qquad
\end{eqnarray}
For example, the two-body $J_{ij}$ interaction can be obtained through linear combinations of the $F_2$ and $F_3$ functions in Eqs.~(\ref{eq:G2-spin}) and~(\ref{eq:G3a-spin}), respectively:
\begin{eqnarray}
	& & J_{ij} = \sum_{\Lambda} \alpha(\Lambda) F_2(R_{ij}; \Lambda) \\
	& & \quad +  \sum_{\Lambda'} \sum_{k \neq i, j} \beta(\Lambda') F_3(R_{ki}, R_{kj}, R_{ij}, \theta_{kij}; \Lambda'). \nonumber
\end{eqnarray}
Here $\alpha$ and $\beta$ are coefficients that are determined through training. 
The first term above, which depends only on the distance $R_{ij}$ between the spin pair, is similar to the RKKY effective interaction discussed in Sec.~\ref{sec:dynamics}. The second $F_3$ term describes a two-spin interaction which depends not only on the pair distance $R_{ij}$, but also on the immediate atomic environment of the spin pair. The $F_3$-type interactions are thus the magnetic analogs of the bond-order potentials, a class of empirical MD potentials that include effects of atomic environment on a chemical bond~\cite{tersoff88,brenner90,pettifor99}. 
The higher-order terms in Eq.~(\ref{eq:E_expansion}) are generated in nonlinear transformations of the neural net. In particular, the four-spin interactions $K_{ijkl}$ are known to play a crucial role in stabilizing non-coplanar spin structures in itinerant electron magnets~\cite{akagi12,hayami17}

Finally, the three-body $F_3$ functions can also be combined with the scalar spin chirality $\chi_{ijk}$ to form a magnetic symmetry functions centered at the $i$-th spin \begin{eqnarray}
	\label{eq:G3b-spin}
	G^\chi_{3}(\Lambda) = \sum_{jk \neq i} F_{3}\bigl(R_{ij}, R_{ik}, R_{jk},\theta_{ijk}; \Lambda) \, (\mathbf S_i \cdot \mathbf S_j \times \mathbf S_k), \qquad
\end{eqnarray}
Since the scalar chirality $\chi_{ijk}$ is nonzero only when the three spins are non-coplanar, the above $G^\chi_{3}$ variables are also very effective in modeling magnetic structures with non-coplanar spins, similar to the 4-spin terms in Eq.~(\ref{eq:E_expansion}). However, unlike the bond variables, the scalar chirality is a pseudo-scalar, which changes sign $\chi_{ijk} \to -\chi_{ijk}$ under time-reversal transformation, or simultaneous inversion of three spins. As a result, they are not appropriate for ML modeling of time-reversal symmetric magnetic systems, such as most spin glasses. One could still include the chirality ACSFs in the feature variables and use data augmentation, i.e. by adding time-reversed spin structures to training dataset, to approximate the time-reversal symmetry. This would lead to, e.g. 6-spin terms $\sum L^{\,}_{ijklmn} \chi^{\,}_{ijk} \chi^{\,}_{lmn}$ in the effective spin Hamiltonian. However, our numerical experiments show that inclusion of chiral ACSFs does not lead to significant improvement.

%We note that generalizations to take into account the different atom species have also been made~\cite{himanen20}. Moreover, depending on the problems at hand, it might be more convenient to use different $F_2$ and $F_3$ functions, and several variants of these functions have been proposed~\cite{himanen20}. 

\section{Relaxation dynamics of an amorphous s-d model}

\label{sec:results}

We apply the above ML framework to a random s-d model as an example for metallic spin glasses. The Hamiltonian in Eq.~(\ref{eq:H1}) describes a general itinerant magnet with s-d type electron-spin coupling. We note that one can also include short-range Heisenberg type interactions $J_{\rm ex} \, \mathbf S_i \cdot \mathbf S_j$ among localized moments, due to either direct or super-exchange mechanisms, to this Hamiltonian.  Since the short-range interaction can be easily included in LLG dynamics simulations, here we focus on ML modeling for the electron-driven forces. 
The details of the electron-mediated interactions depend on the electronic Hamiltonian $\hat{\mathcal{H}}_e$. For example, the well-studied s-d model, also known as the Kondo-lattice model for spin-1/2, corresponds to a periodic array of local spins $\mathbf S_i$ and a tight-binding Hamiltonian with nearest-neighbor hopping defined on the same lattice~\cite{furukawa94,dagotto98,pekker05}. The ferromagnetic s-d model in the strong coupling limit, also known as the double-exchange model, plays an important role in the colossal magnetoresistance phenomena~\cite{yunoki98,dagotto03}.  

% In the absence of such short-range exchange terms, the effective spin-spin interaction originates from electron-mediated couplings.

Here we introduce an amorphous generalization of the s-d model. First, instead of placing the spins on a periodic lattice, the local moments are randomly distributed in a three dimensional space. Specifically, the position $\mathbf r_i$ of spin $\mathbf S_i$ is a random vector uniformly distributed within a 3D cubical box of side $L$, with the only constraint that the distance between any pair of spins is greater than a minimum, i.e. $R_{ij} = |\mathbf r_j - \mathbf r_i| > r_{\rm min}$. For a given random atomic configuration $\{\mathbf r_i\}$, a disordered tight-binding model is employed to describe the electronic subsystem $\hat{\mathcal{H}}_e$, giving rise to a disordered s-d Hamiltonian
\begin{eqnarray}
	\label{eq:H_sd}
	& & \hat{\mathcal{H}} = \sum_{ij}\sum_{\alpha = \uparrow, \downarrow} t\bigl(|\mathbf r_i - \mathbf r_j| \bigr) \, c^\dagger_{i \alpha} c^{\,}_{j \alpha} \nonumber \\
	& & \qquad - J \sum_i \sum_{\alpha,\beta = \uparrow,\downarrow} \mathbf S_i \cdot   \left(\hat{c}^\dagger_{i, \alpha} \bm\sigma^{\,}_{\alpha\beta} \hat{c}^{\,}_{i, \beta} \right). 
\end{eqnarray}
The electron hopping coefficient $t_{ij} = t(R_{ij})$ is a random variable dependent on the distance $R_{ij}$ between an atomic pair through a Yukawa-type exponentially decaying function 
\begin{eqnarray}
	t(R) = t_0 \exp(-R / \ell). 
\end{eqnarray}
In the following, we set $t_0 = 1$ which also serves as the reference for energy. The characteristic range or length scale of electron hopping is given by the decay length~$\ell$. It is worth noting that such Yukawa tight-binding models have long been used in the modeling of amorphous systems~\cite{ching82,logan88,winn89,bush89,priour12}.

\begin{figure}[t]
\centering
\includegraphics[width=0.99\columnwidth]{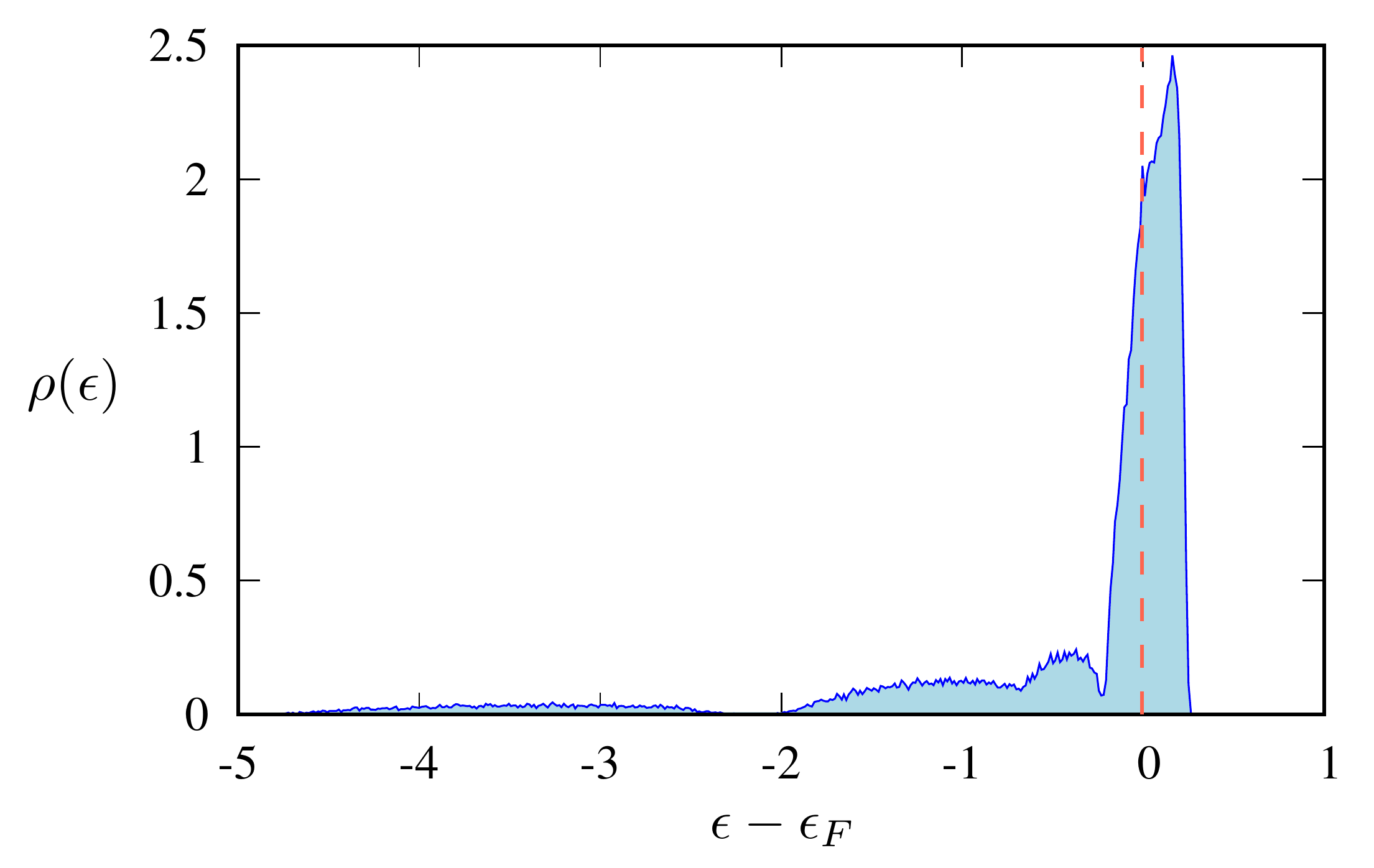}
\caption{Electron density of states of the disordered tight-binding model Eq.~(\ref{eq:H_sd}) in the absence of electron-spin coupling $J = 0$. The hopping constant $t_0 = 1$, which serves as the unit of the energy. The number of atoms is $N = 100$ in a cubical box of linear size $L = 5 \ell$ with periodic boundary conditions. The electron filling fraction is set at $f = 0.5$.  The zero of the energy, indicated by the dashed line, corresponds to the Fermi level. }
    \label{fig:dos}
\end{figure}

We first examine the electron density of states (DOS) of the random tight-binding Hamiltonian. To this end, we set the s-d coupling $J = 0$ and use exact diagonalization to compute the eigen-energies $\epsilon_k$ of the electron hopping Hamiltonian. The DOS, defined as $\rho_e(\epsilon) = \frac{1}{N} \langle \sum_k \delta(\epsilon - \epsilon_k) \rangle$, is obtained by averaging 500 different atomic configurations $\{\mathbf r_i\}$. In this calculation, $N = 500$ atoms are randomly distributed within a simulation box of linear size $L = 5 \ell$, with a minimum separation $r_{\rm min} = 0.5 \ell$. The resultant DOS is shown in FIG.~\ref{fig:dos}. The Fermi energy $\epsilon_F$ is determined from the condition of half electron filling. The DOS exhibits a pronounced peak in the vicinity of the Fermi level. The nonzero DOS at Fermi level $\rho(\epsilon_F)$ indicates a gapless electronic state that is susceptible to small perturbations. Moreover, previous large-scale numerical study shows that electron wave functions for the eigenstates in the middle of the band is likely to be delocalized~\cite{priour12}. These extended electron eigenstates near the Fermi level are the dominate contributors to the long-range effective spin-spin interactions. 

Our goal is to build a ML model for the disordered s-d system~(\ref{eq:H_sd}) with a large electron-spin coupling $J = 6 t_0$. We note that the RKKY type perturbation methods cannot be applied to such strong coupling regime of s-d type models. The ML methods thus provide a non-perturbative approach to derive an effective classical spin Hamiltonian and force field that are beyond the analytical methods. 
To generate the training and testing datasets, ED and ED-LLG simulations are carried out on a system consisting of $N = 100$ atoms with a half-filled electron band in a cubical box of linear size $L = 5 \ell$. The minimum separation is again set at $r_{\rm min} = 0.5 \ell$.  10~different realizations of random atomic configurations are used in generating the dataset. For each realization of the atomic structure, the s-d model is solved using the ED for 250 different spin configurations, including random spins and states obtained from relaxation simulations. It is worth noting that, as the prediction of local field $\mathbf H_i$ for each spin counts as single training data point, the size of the effective datasets is $250 \times 10 \times 100 = 2.5\times 10^5$.

A neural network with 8 hidden layers is constructed using PyTorch~\cite{paszke19} to learn the dependence of local energy $\epsilon_i$ on the feature variables $\{G_k\}$ that characterize the neighborhood $\mathcal{C}_i$. The number of neurons at the input layer is determined by the number of feature variables and is fixed at 450. The number of neurons in successive hidden layers are: 1024 $\times$ 1024 $\times$ 512 $\times$ 256 $\times$ 128 $\times$ 128 $\times$ 128 $\times$ 128. The NN performs a series of nonlinear transformations on the input neurons where ReLU~\cite{barron17} is used as the activation function between layers. The NN model is trained based on a loss function including the mean square error (MSE) of both the effective field and total energy. Since only the perpendicular component of the local field $\mathbf H_{i, \perp} = \mathbf S_i \times \mathbf H_i$ contributes to the driving force for spin dynamics, the loss function focuses on the MSE of the perpendicular field. Specifically, for a given spin configuration, the loss function is defined as
\begin{eqnarray}
	L = \mu_H \sum_{i=1}^N \Bigl| \mathbf H^{\rm ED}_{i, \perp}  - \mathbf H^{\rm ML}_{i, \perp}   \Bigr|^2 
	+ \mu_E \Bigl| E^{\rm ED} - E^{\rm ML} \Bigr|^2, \quad
\end{eqnarray}
where $\mu_H$ and $\mu_E$ determines the relative weights of the force and energy constraints in the loss function. As shown in Eq.~(\ref{eq:H_ML}), the ML local field is obtained from the derivative of the sum of local energies. This can be efficiently done using automatic differentiation in PyTorch~\cite{paszke17}. Trainable parameters of the NN are optimized by the Adam stochastic optimizer~\cite{kingma17} with a learning rate of 0.001. A 5-fold cross-validation and early stopping regularization are performed to prevent overfitting.

\begin{figure}
\centering
\includegraphics[width=0.99\columnwidth]{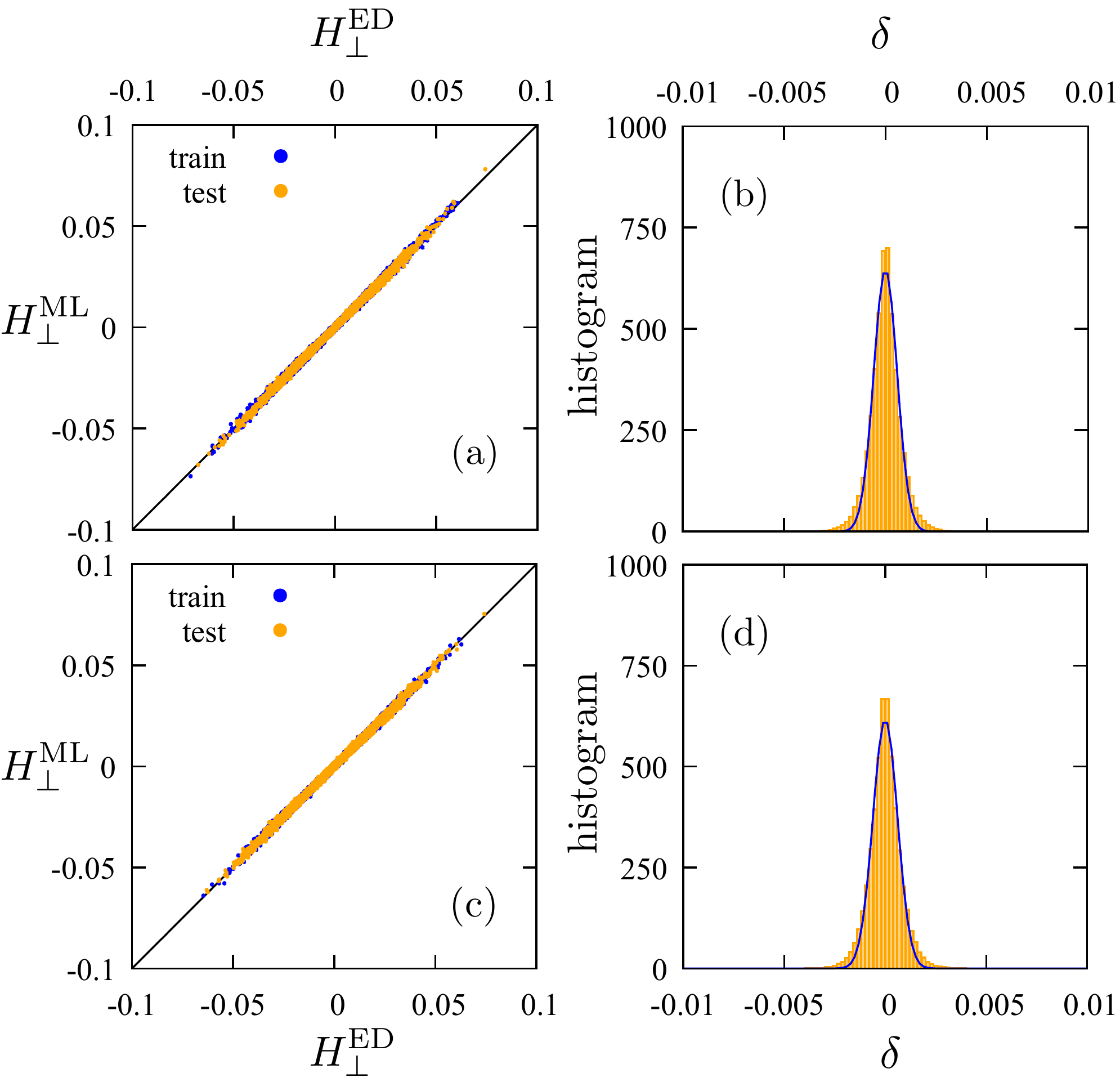}
\caption{Benchmark of ML prediction for the perpendicular local field. Panel~(a) shows the predicted components of $\mathbf H_\perp$ versus the ground-truth values for the trained NN model based on a magnetic descriptor with all three symmetry functions $G^m_2$, $G^m_3$, and $G^\chi_3$. The corresponding histogram of prediction error $\delta = H^{\rm ML}_\perp - H^{\rm ED}_\perp$ is shown in panel~(b). The results for ML model with a descriptor that {\em excludes} the chirality feature variables $G^\chi_3$ are shown in panels~(c) and (d). }
    \label{fig:force}
\end{figure}

To incorporate the rotation symmetry, both in real and spin space, to the ML model, a descriptor with all three $G^m_2$, $G^m_3$ and $G^\chi_3$ symmetry functions is developed to characterize local spin configurations $\mathcal{C}_i$. A cutoff radius $R_c = 2.5 \ell$ is used for computing these feature variables. First, the two-body symmetry functions $G^m_2(d, w)$ depend on two parameters characterizing a mass shell of thickness $w$ and radius $d$. The width of the shell is fixed at $w = 0.05\ell$, while 50 different radii $d$ in the range of $[0.5, 2.46]$ (in units of $\ell$) are used. 
For both three-body symmetry functions $G^m_3$ and $G^\chi_3$, there are two parameters $\zeta, \lambda$ for characterizing the angular distribution, similar $d, w$ for a mass shell, and another set $d', w'$ for the distance between two neighboring spins; see Eq.~(\ref{eq:F3-envelop}). Here we set $\zeta = 1$ and consider two $\lambda = \pm 1$ corresponding to an angular function $(1 \pm \cos\theta_{ijk})$ with  a peak at a relative angle $\theta_{ijk} = 0$ and $\pi$, respectively. For the three $F_2$ functions in $G^m_3$ and $G^\chi_3$, we use $w' = 1$ and two different $d' = 1$ and 2 in units of $\ell$, and the same $w, d$ parameters as in the $G^m_2$. The total number of feature variables is $50 + 2 \times 200 = 450$.

The benchmark of force prediction for the above ML model with all three $G^m_2$, $G^m_3$, and $G^\chi_3$ symmetry functions is shown in FIG.~\ref{fig:force}(a) and (b). Excellent agreements between the ML predicted local fields and the exact values are obtained for both training and testing datasets. The histogram of the prediction error, shown in FIG.~\ref{fig:force}(b), is characterized by a rather small mean square error $\sigma \approx 0.001$. However, as discussed in Sec.~\ref{sec:descriptor}, the chirality feature variables $G^\chi_3$ change sign under the time-reversal transformation. The resultant effective spin Hamiltonian and the force-field model is not invariant with respect to time-reversal symmetry, which is an intrinsic symmetry of the original s-d model. The time-reversal symmetry can be incorporated into the ML model, albeit inexactly, using data augmentation, i.e. by including both spin configurations $\pm \mathbf S_i$ in the dataset. On the other hand, effective interactions involving scalar chirality that is also time-reversal symmetric, $\sum L_{ijklmn} \chi_{ijk} \chi_{lmn}$, is of six-order in spin variables. It is likely that their contribution is negligible compared to that of bond variables.

For comparison, we have also developed a ML model with a magnetic descriptor only involves the two magnetic symmetry functions $G^m_2$ and $G^m_3$. Since these two feature variables are built from bond variables $b_{ij} = \mathbf S_i \cdot \mathbf S_j$, hence are invariant under time-reversal, the resultant ML models automatically preserve the time-reversal symmetry. The force benchmark of this ML model {\em without} chirality variables is summarized in FIG.~\ref{fig:force}(c) and~(d). Notably, the ML predicted local fields also agree very well with the ED calculations. This result indicates that the spin chirality variables do not play a major role in the effective spin Hamiltonian. The bond variables alone provide accurate approximations to the amorphous s-d model. 

\begin{figure}
\centering
\includegraphics[width=0.99\columnwidth]{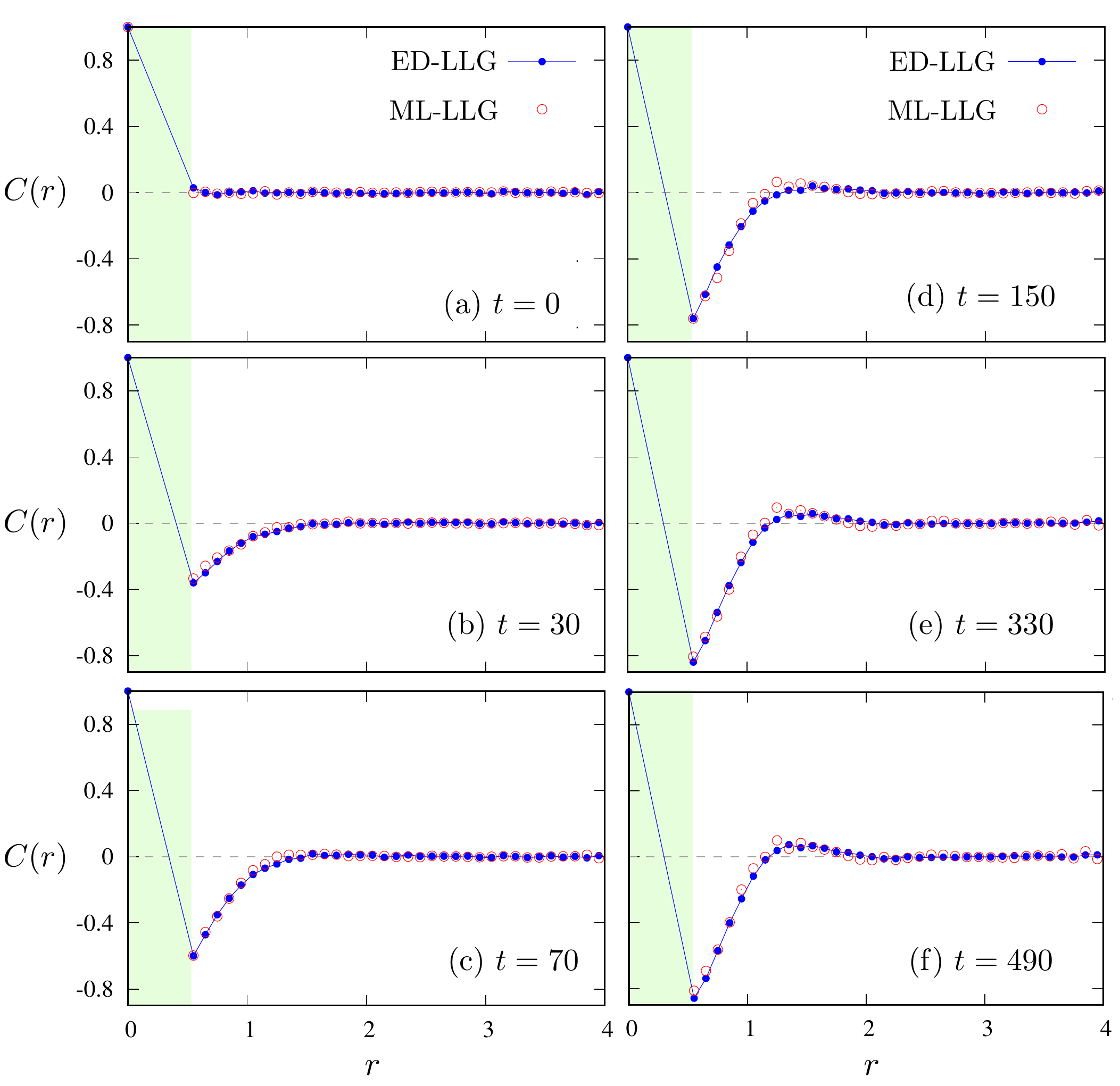}
\caption{Dynamical benchmark of the ML force-field model for the random s-d system. The spin-spin correlation functions $C(r_{ij}) = \langle \mathbf S_i \cdot \mathbf S_j\rangle$ at various time-steps after a thermal quench are obtained from LLG simulations of $N = 100$ spins using the ML force-field model and the ED method. The radius is measured in units of $\ell$, and the shaded area indicates the forbidden region where $r < r_{\rm min}$. The simulation time is measured in units of $(\gamma t_0)^{-1}$, where $\gamma$ is the gyromagnetic ratio and $t_0$ is the energy scale of electron hopping.}
    \label{fig:corr-t}
\end{figure}

Next we perform dynamical benchmarks of the ML force-field models. To this end, we integrate the trained ML models with the LLG method and perform thermal-quench simulations of the random s-d model. Specifically, for a given disordered atomic configuration, an initial state of random spins is quenched to a temperature $T = 0.001 t_0$ at time $t = 0$. A damping coefficient $\alpha = 0.05 \gamma$ in Eq.~(\ref{eq:LLG}), where $\gamma$ is the gyromagnetic ratio, is used for both the ML- and ED-LLG simulations. The LLG simulations are then repeated for different independent initial random spins and realization of atomic disorder. The ensemble-averaged dynamical evolution of the quenched states is then compared with that obtained from the ED-LLG simulations.  In particular, we compute the time-dependent correlation functions 
\begin{eqnarray}
	C(r_{ij}, t) = \langle \mathbf S_i(t) \cdot \mathbf S_j(t) \rangle - \langle \mathbf S_i(t) \rangle \cdot \langle \mathbf S_j(t) \rangle,
\end{eqnarray}
where $\langle \cdots \rangle$ means ensemble average over both atomic and spin configurations. FIG.~\ref{fig:corr-t} shows the ensemble-averaged spin-spin correlation functions at different times during the quench. The correlations obtained from the ML-LLG simulations agree well with those from the ED methods, showing that the ML force-field model not only accurately predicts the driving forces, but also capture the spin dynamics and relaxation process.  
 
The correlation functions are assumed to depend only on the distance $r_{ij}$ between of a pair of spins, thanks to the rotation and translation symmetry of the disordered states. 
The ensemble-averaged magnetization is found to be nearly zero, $\mathbf m = \langle \mathbf S_i \rangle \approx \mathbf 0$, in both ML and ED-LLG simulations. Moreover, we find that the relaxation dynamics slows down significantly at large times after the quench. The freezing of spin dynamics thus indicates that the system settles into a local minimum of the effective spin Hamiltonian.

Due to the minimum distance constraint, there are no spin pairs with distance less than $r_{\rm min} = 0.5 \ell$. On the other hand, early in the relaxation, e.g. at $t \lesssim 30$ in units of $(\gamma t_0)^{-1}$,  an anti-parallel spin-spin correlation $C(r) \approx -0.4$ quickly develops at the minimum separation~$r_{\rm min}$. This negative correlation then decays with increasing separation and vanishes at $r \approx 1.4\ell$. This result shows that the effective spin-spin interaction is predominantly antiferromagnetic at short distances. Indeed, as has been demonstrated in the case of lattice s-d models at half-filling, the electron-mediated interaction in the large coupling limit is antiferromagnetic~\cite{chern18}. This can be understood from the $J \to \infty$  (or $t_0 = 0$) limit where the ground states are given by configurations with exactly one electron per atom. As atoms are decoupled from each other, the local moment $\mathbf S_i$, which is entangled with the spin of localized electron, can point in arbitrary directions. This macroscopic ground-state degeneracy is lifted in the presence of a small $t_0$. Standard second-order perturbation leads to an effective antiferromagnetic interaction $J^{\rm eff}_{ij} = t_{ij}^2 / J > 0$.

As the system further relaxes toward lower energy states, a weak ferromagnetic correlation starts to develop at an intermediate distance, as indicated by the small peak at $r \sim 1.5\ell$ for the correlation functions at $t \gtrsim 150$. Compared with lattice models where electron hopping is restricted to nearest neighbors, the emergence of this ferromagnetic correlation at intermediate separation indicates the complexity of electron-mediated interactions in disordered s-d systems. Interestingly, our results show that the effective spin-spin interactions exhibits an RKKY-like oscillation, although with a much reduced amplitude, even in the large coupling regime. 

\begin{figure}
\centering
\includegraphics[width=0.9\columnwidth]{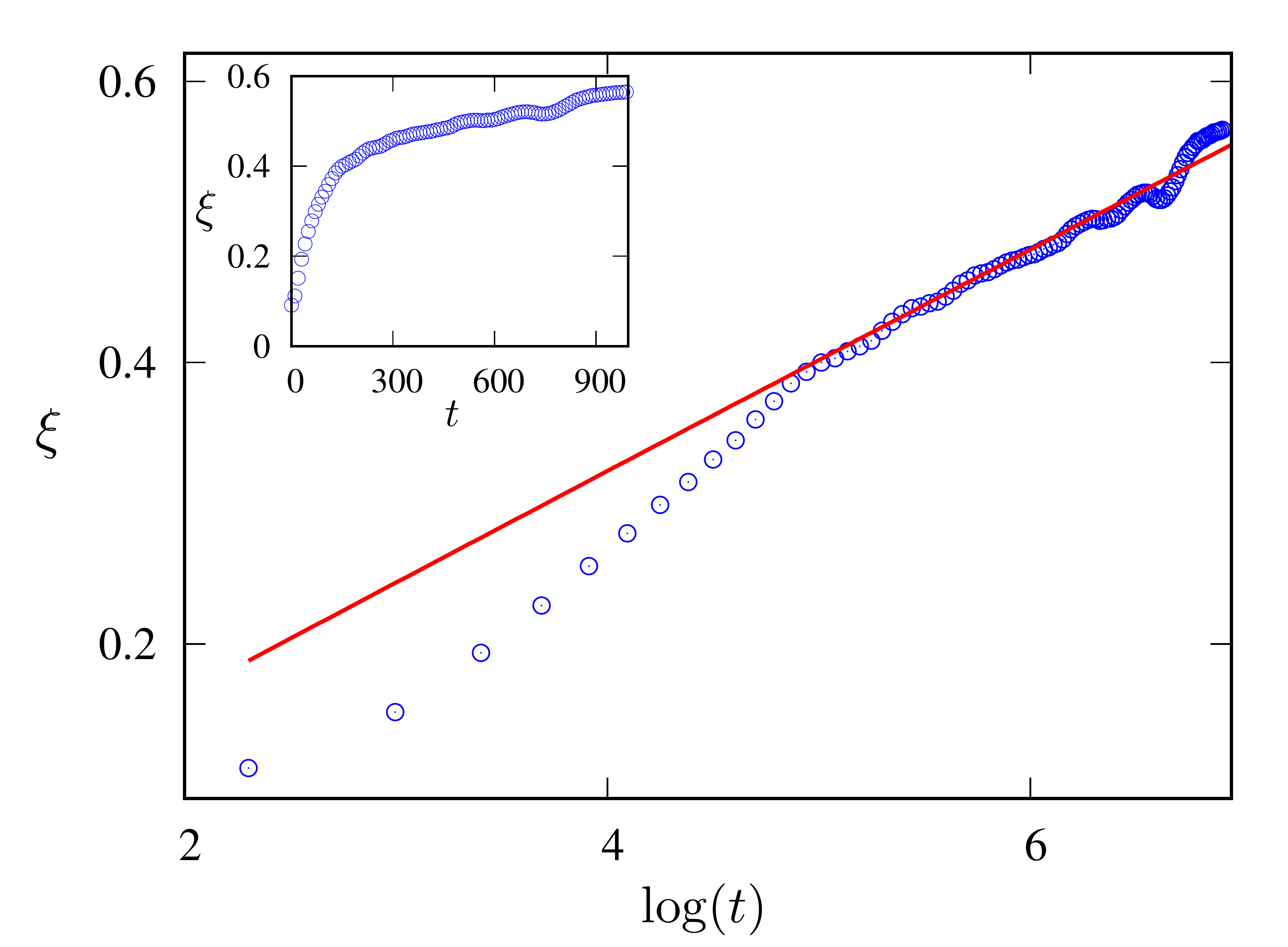}
\caption{The correlation length $\xi$ as a  function of the the logarithm of time $\log(t)$ in a log-log plot; the inset shows the $\xi(t)$ in the linear scale. ML-LLG simulations are performed on a random s-d system of $N = 500$ atoms. The length $\xi$ is computed from the resultant ensemble-averaged correlation functions after a thermal quench at $t = 0$. The simulation time is measured in units of $(\gamma t_0)^{-1}$, where $\gamma$ is the gyromagnetic ratio and $t_0$ is the energy scale of electron hopping. The red line corresponds to a power law $\xi(t) \sim (\log t)^{1/\psi}$ with an exponent $\psi \approx 1.02$.}
    \label{fig:corr-Lt}
\end{figure}

To quantify the relaxation process, we compute a time-dependent correlation length $\xi(t)$ from the correlation functions. For a well-defined exponential decaying correlation, $C(r) \sim \exp(-r/\xi)$, the correlation length can be easily computed from the large $r$ behavior. Due to the more complicated forms of the spin-spin correlation functions in our case, we employ an empirical formula to obtain the correlation length $\xi(t) = \int_{0}^\infty r \bigl| C(r, t) \bigr| dr \Big/ \int_{0}^\infty \bigl| C(r, t) \bigr| dr$,
It can be easily shown that this formula correctly reproduces the correlation length for an exponential-decaying function as well as for correlation functions exhibiting a dynamical scaling: $C(r, t) = \mathcal{F}(r/\xi(t))$, where $\mathcal{F}(x)$ is the scaling function. Applying this formula to the correlation functions obtained from ML-LLG simulations for a random s-d model with $N = 500$ atoms, the result is shown in the inset of Fig.~\ref{fig:corr-Lt}. The ML model is trained from a system of $N = 100$ atoms with the same model parameters as discussed above. The ensemble-averaged correlation functions here are obtained from 10 different random realizations of atomic configurations and 20 independent initial random spin states. 

As the system relaxes after the thermal quench, the correlation length quickly increases initially. The late-stage relaxation dynamics of a spin glass, often characterized by aging phenomena, is intimately related to its complex energy landscape. Although there is no conventional long-range order in spin glasses, the relaxation processes can still be viewed as the growth of ``ordered" domains in which a local energy minimum is attained. It is worth noting that growth of ordered domains in a conventional symmetry breaking phase is controlled by topological defects of the associated order parameter fields and is often characterized by a power-law behavior~\cite{bray94}. 

On the other hand, the growth of the spin-glass order has been discussed within the phenomenological droplet model, which predicts a growth law $R(t) \sim (\log t)^{1/\psi}$, where $R(t)$ is a measure of the linear domain size and $\psi$ is called the barrier exponent which characterizes the activation barrier height at the interface~\cite{fisher88,huse91}. Using the extracted correlation length as a qualitative measure of the domain size, Fig.~\ref{fig:corr-Lt} shows $\xi(t)$ versus $\log(t)$ in a log-log plot. The late-stage domain coarsening can be approximated by a power-law with an exponent $\psi \approx 1.02$. Our preliminary results are consistent with the scaling theory of the droplet model. More systematic investigation of the aging dynamics and domain growth will be left for future work.

\section{Conclusion and outlook}

\label{sec:conclusion}

To summarize, we have proposed a scalable ML force-field model for spin dynamics in metallic spin glasses. Our approach is a generalization of the Behler-Parrinello schemes which are widely used in ML-based force-field models for quantum molecular dynamics. We focus on spin glass systems with Heisenberg spins in three dimensions, which are also the case for prototype metallic spin-glass materials such as dilute magnetic alloys. To preserve the rotation symmetry in both real and spin spaces, we have generalized the ACSF atomic descriptor, which is widely used in ML-based MD simulations, by incorporating spin degrees of freedom into the symmetry functions. As a demonstration, we apply our approach to study the relaxation dynamics of an amorphous generalization of the well-known s-d model. We show that the trained ML model not only accurately predicts the effective local fields, but also captures the dynamical evolution of spins under a thermal quench. 

Our work also opens a new avenue to large-scale simulations of nonequilibrium dynamical phenomena in spin glasses. Partly because of the computational complexity, most large-scale studies of spin-glass dynamics are based on short-range random-$J$ spin models on a lattice. Moreover, local Monte Carlo updates are employed as a surrogate dynamics for Ising-type spin-glass models. Here we outline a general approach to derive realistic off-lattice spin-glass model with electron-mediated spin-spin interactions. Our work demonstrates the proof of principle that accurate ML force-field models can be developed for a large class of metallic spin glass systems with local s-d electron-spin coupling. Importantly, the efficiency of ML models, which are essentially effective classical spin models, could enable large-scale simulations which are essential for understanding the dynamical properties of spin glass systems.

In particular, our ML framework could also be used to model dilute magnetic alloys such as CuMn. As discussed in the Introduction, DFT has been applied to solve the magnetic ground states of a few representative magnetic alloys for a given atomic configuration where the magnetic atoms randomly occupy sites of the host lattice. However, such first-principles calculations of the magnetic moments and the corresponding effective fields are very time-consuming even for relatively small systems. In practice, often an effective Heisenberg model is first obtained by fitting the exchange interactions with DFT energy. An alternative approach is to derive an effective s-d Hamiltonian from first-principles calculations, which can then be used to develop a ML force field model as outlined in this work. 

%Finally, while this work focuses on ML force-field models for spin dynamics, the proposed framework can also be straightforwardly to include atomic forces for molecular dynamics simulations. 

\begin{acknowledgments}
We thank useful discussions with Yunhao Fan and Kotaro Shimizu. This work was supported by the US Department of Energy Basic Energy Sciences under Award No. DE-SC0020330. The authors also acknowledge the support of Research Computing at the University of Virginia.
\end{acknowledgments}

\bibliography{ref}

%apsrev4-2.bst 2019-01-14 (MD) hand-edited version of apsrev4-1.bst
%Control: key (0)
%Control: author (8) initials jnrlst
%Control: editor formatted (1) identically to author
%Control: production of article title (0) allowed
%Control: page (0) single
%Control: year (1) truncated
%Control: production of eprint (0) enabled
\begin{thebibliography}{86}%
\makeatletter
\providecommand \@ifxundefined [1]{%
 \@ifx{#1\undefined}
}%
\providecommand \@ifnum [1]{%
 \ifnum #1\expandafter \@firstoftwo
 \else \expandafter \@secondoftwo
 \fi
}%
\providecommand \@ifx [1]{%
 \ifx #1\expandafter \@firstoftwo
 \else \expandafter \@secondoftwo
 \fi
}%
\providecommand \natexlab [1]{#1}%
\providecommand \enquote  [1]{``#1''}%
\providecommand \bibnamefont  [1]{#1}%
\providecommand \bibfnamefont [1]{#1}%
\providecommand \citenamefont [1]{#1}%
\providecommand \href@noop [0]{\@secondoftwo}%
\providecommand \href [0]{\begingroup \@sanitize@url \@href}%
\providecommand \@href[1]{\@@startlink{#1}\@@href}%
\providecommand \@@href[1]{\endgroup#1\@@endlink}%
\providecommand \@sanitize@url [0]{\catcode `\\12\catcode `\$12\catcode
  `\&12\catcode `\#12\catcode `\^12\catcode `\_12\catcode `\%12\relax}%
\providecommand \@@startlink[1]{}%
\providecommand \@@endlink[0]{}%
\providecommand \url  [0]{\begingroup\@sanitize@url \@url }%
\providecommand \@url [1]{\endgroup\@href {#1}{\urlprefix }}%
\providecommand \urlprefix  [0]{URL }%
\providecommand \Eprint [0]{\href }%
\providecommand \doibase [0]{https://doi.org/}%
\providecommand \selectlanguage [0]{\@gobble}%
\providecommand \bibinfo  [0]{\@secondoftwo}%
\providecommand \bibfield  [0]{\@secondoftwo}%
\providecommand \translation [1]{[#1]}%
\providecommand \BibitemOpen [0]{}%
\providecommand \bibitemStop [0]{}%
\providecommand \bibitemNoStop [0]{.\EOS\space}%
\providecommand \EOS [0]{\spacefactor3000\relax}%
\providecommand \BibitemShut  [1]{\csname bibitem#1\endcsname}%
\let\auto@bib@innerbib\@empty
%</preamble>
\bibitem [{\citenamefont {Mezard}\ and\ \citenamefont
  {Virasoro}(1987)}]{mezard87}%
  \BibitemOpen
  \bibfield  {author} {\bibinfo {author} {\bibfnamefont {P.~G.}\ \bibnamefont
  {Mezard}, \bibfnamefont {M.}}\ and\ \bibinfo {author} {\bibfnamefont
  {M.}~\bibnamefont {Virasoro}},\ }\href@noop {} {\emph {\bibinfo {title} {Spin
  Glass Theory and Beyond}}}\ (\bibinfo  {publisher} {World Scientific},\
  \bibinfo {address} {Singapore},\ \bibinfo {year} {1987})\BibitemShut
  {NoStop}%
\bibitem [{\citenamefont {Binder}\ and\ \citenamefont
  {Young}(1986)}]{binder86}%
  \BibitemOpen
  \bibfield  {author} {\bibinfo {author} {\bibfnamefont {K.}~\bibnamefont
  {Binder}}\ and\ \bibinfo {author} {\bibfnamefont {A.~P.}\ \bibnamefont
  {Young}},\ }\bibfield  {title} {\bibinfo {title} {Spin glasses: Experimental
  facts, theoretical concepts, and open questions},\ }\href
  {https://doi.org/10.1103/RevModPhys.58.801} {\bibfield  {journal} {\bibinfo
  {journal} {Rev. Mod. Phys.}\ }\textbf {\bibinfo {volume} {58}},\ \bibinfo
  {pages} {801} (\bibinfo {year} {1986})}\BibitemShut {NoStop}%
\bibitem [{\citenamefont {Mydosh}(1993)}]{mydosh95}%
  \BibitemOpen
  \bibfield  {author} {\bibinfo {author} {\bibfnamefont {J.~A.}\ \bibnamefont
  {Mydosh}},\ }\href@noop {} {\emph {\bibinfo {title} {Spin Glasses: An
  Experimental Introduction}}}\ (\bibinfo  {publisher} {Taylor \& Francis},\
  \bibinfo {address} {London},\ \bibinfo {year} {1993})\BibitemShut {NoStop}%
\bibitem [{\citenamefont {Mydosh}(2015)}]{mydosh15}%
  \BibitemOpen
  \bibfield  {author} {\bibinfo {author} {\bibfnamefont {J.~A.}\ \bibnamefont
  {Mydosh}},\ }\bibfield  {title} {\bibinfo {title} {Spin glasses: redux: an
  updated experimental/materials survey},\ }\href
  {https://doi.org/10.1088/0034-4885/78/5/052501} {\bibfield  {journal}
  {\bibinfo  {journal} {Reports on Progress in Physics}\ }\textbf {\bibinfo
  {volume} {78}},\ \bibinfo {pages} {052501} (\bibinfo {year}
  {2015})}\BibitemShut {NoStop}%
\bibitem [{\citenamefont {Dyre}(2006)}]{dyre06}%
  \BibitemOpen
  \bibfield  {author} {\bibinfo {author} {\bibfnamefont {J.~C.}\ \bibnamefont
  {Dyre}},\ }\bibfield  {title} {\bibinfo {title} {Colloquium: The glass
  transition and elastic models of glass-forming liquids},\ }\href
  {https://doi.org/10.1103/RevModPhys.78.953} {\bibfield  {journal} {\bibinfo
  {journal} {Rev. Mod. Phys.}\ }\textbf {\bibinfo {volume} {78}},\ \bibinfo
  {pages} {953} (\bibinfo {year} {2006})}\BibitemShut {NoStop}%
\bibitem [{\citenamefont {Charbonneau}\ \emph {et~al.}(2017)\citenamefont
  {Charbonneau}, \citenamefont {Kurchan}, \citenamefont {Parisi}, \citenamefont
  {Urbani},\ and\ \citenamefont {Zamponi}}]{charbonneau17}%
  \BibitemOpen
  \bibfield  {author} {\bibinfo {author} {\bibfnamefont {P.}~\bibnamefont
  {Charbonneau}}, \bibinfo {author} {\bibfnamefont {J.}~\bibnamefont
  {Kurchan}}, \bibinfo {author} {\bibfnamefont {G.}~\bibnamefont {Parisi}},
  \bibinfo {author} {\bibfnamefont {P.}~\bibnamefont {Urbani}},\ and\ \bibinfo
  {author} {\bibfnamefont {F.}~\bibnamefont {Zamponi}},\ }\bibfield  {title}
  {\bibinfo {title} {Glass and jamming transitions: From exact results to
  finite-dimensional descriptions},\ }\href
  {https://doi.org/10.1146/annurev-conmatphys-031016-025334} {\bibfield
  {journal} {\bibinfo  {journal} {Annual Review of Condensed Matter Physics}\
  }\textbf {\bibinfo {volume} {8}},\ \bibinfo {pages} {265} (\bibinfo {year}
  {2017})}\BibitemShut {NoStop}%
\bibitem [{\citenamefont {Edwards}\ and\ \citenamefont
  {Anderson}(1975)}]{edwards75}%
  \BibitemOpen
  \bibfield  {author} {\bibinfo {author} {\bibfnamefont {S.~F.}\ \bibnamefont
  {Edwards}}\ and\ \bibinfo {author} {\bibfnamefont {P.~W.}\ \bibnamefont
  {Anderson}},\ }\bibfield  {title} {\bibinfo {title} {Theory of spin
  glasses},\ }\href {https://doi.org/10.1088/0305-4608/5/5/017} {\bibfield
  {journal} {\bibinfo  {journal} {Journal of Physics F: Metal Physics}\
  }\textbf {\bibinfo {volume} {5}},\ \bibinfo {pages} {965} (\bibinfo {year}
  {1975})}\BibitemShut {NoStop}%
\bibitem [{\citenamefont {Sherrington}\ and\ \citenamefont
  {Kirkpatrick}(1975)}]{sherrington75}%
  \BibitemOpen
  \bibfield  {author} {\bibinfo {author} {\bibfnamefont {D.}~\bibnamefont
  {Sherrington}}\ and\ \bibinfo {author} {\bibfnamefont {S.}~\bibnamefont
  {Kirkpatrick}},\ }\bibfield  {title} {\bibinfo {title} {Solvable model of a
  spin-glass},\ }\href {https://doi.org/10.1103/PhysRevLett.35.1792} {\bibfield
   {journal} {\bibinfo  {journal} {Phys. Rev. Lett.}\ }\textbf {\bibinfo
  {volume} {35}},\ \bibinfo {pages} {1792} (\bibinfo {year}
  {1975})}\BibitemShut {NoStop}%
\bibitem [{\citenamefont {D.~J.~Thouless}\ and\ \citenamefont
  {Palmer}(1977)}]{thouless77}%
  \BibitemOpen
  \bibfield  {author} {\bibinfo {author} {\bibfnamefont {P.~W.~A.}\
  \bibnamefont {D.~J.~Thouless}}\ and\ \bibinfo {author} {\bibfnamefont
  {R.~G.}\ \bibnamefont {Palmer}},\ }\bibfield  {title} {\bibinfo {title}
  {Solution of 'solvable model of a spin glass'},\ }\href
  {https://doi.org/10.1080/14786437708235992} {\bibfield  {journal} {\bibinfo
  {journal} {The Philosophical Magazine: A Journal of Theoretical Experimental
  and Applied Physics}\ }\textbf {\bibinfo {volume} {35}},\ \bibinfo {pages}
  {593} (\bibinfo {year} {1977})}\BibitemShut {NoStop}%
\bibitem [{\citenamefont {Parisi}(1979)}]{parisi79}%
  \BibitemOpen
  \bibfield  {author} {\bibinfo {author} {\bibfnamefont {G.}~\bibnamefont
  {Parisi}},\ }\bibfield  {title} {\bibinfo {title} {Infinite number of order
  parameters for spin-glasses},\ }\href
  {https://doi.org/10.1103/PhysRevLett.43.1754} {\bibfield  {journal} {\bibinfo
   {journal} {Phys. Rev. Lett.}\ }\textbf {\bibinfo {volume} {43}},\ \bibinfo
  {pages} {1754} (\bibinfo {year} {1979})}\BibitemShut {NoStop}%
\bibitem [{\citenamefont {Parisi}(1983)}]{parisi83}%
  \BibitemOpen
  \bibfield  {author} {\bibinfo {author} {\bibfnamefont {G.}~\bibnamefont
  {Parisi}},\ }\bibfield  {title} {\bibinfo {title} {Order parameter for
  spin-glasses},\ }\href {https://doi.org/10.1103/PhysRevLett.50.1946}
  {\bibfield  {journal} {\bibinfo  {journal} {Phys. Rev. Lett.}\ }\textbf
  {\bibinfo {volume} {50}},\ \bibinfo {pages} {1946} (\bibinfo {year}
  {1983})}\BibitemShut {NoStop}%
\bibitem [{\citenamefont {M\'ezard}\ \emph {et~al.}(1984)\citenamefont
  {M\'ezard}, \citenamefont {Parisi}, \citenamefont {Sourlas}, \citenamefont
  {Toulouse},\ and\ \citenamefont {Virasoro}}]{mezard84}%
  \BibitemOpen
  \bibfield  {author} {\bibinfo {author} {\bibfnamefont {M.}~\bibnamefont
  {M\'ezard}}, \bibinfo {author} {\bibfnamefont {G.}~\bibnamefont {Parisi}},
  \bibinfo {author} {\bibfnamefont {N.}~\bibnamefont {Sourlas}}, \bibinfo
  {author} {\bibfnamefont {G.}~\bibnamefont {Toulouse}},\ and\ \bibinfo
  {author} {\bibfnamefont {M.}~\bibnamefont {Virasoro}},\ }\bibfield  {title}
  {\bibinfo {title} {Nature of the spin-glass phase},\ }\href
  {https://doi.org/10.1103/PhysRevLett.52.1156} {\bibfield  {journal} {\bibinfo
   {journal} {Phys. Rev. Lett.}\ }\textbf {\bibinfo {volume} {52}},\ \bibinfo
  {pages} {1156} (\bibinfo {year} {1984})}\BibitemShut {NoStop}%
\bibitem [{\citenamefont {Nishimori}(2001)}]{nishimori01}%
  \BibitemOpen
  \bibfield  {author} {\bibinfo {author} {\bibfnamefont {H.}~\bibnamefont
  {Nishimori}},\ }\href@noop {} {\emph {\bibinfo {title} {Statistical Physics
  of Spin Glasses and Information Processing}}}\ (\bibinfo  {publisher} {Oxford
  University Press},\ \bibinfo {year} {2001})\BibitemShut {NoStop}%
\bibitem [{\citenamefont {Stein}\ and\ \citenamefont {Newman}(2013)}]{stein13}%
  \BibitemOpen
  \bibfield  {author} {\bibinfo {author} {\bibfnamefont {D.~L.}\ \bibnamefont
  {Stein}}\ and\ \bibinfo {author} {\bibfnamefont {C.~M.}\ \bibnamefont
  {Newman}},\ }\href@noop {} {\emph {\bibinfo {title} {Spin Glasses and
  Complexity}}}\ (\bibinfo  {publisher} {Princeton University Press},\ \bibinfo
  {year} {2013})\BibitemShut {NoStop}%
\bibitem [{\citenamefont {Charbonneau}(2023)}]{charbonneau23}%
  \BibitemOpen
  \bibinfo {editor} {\bibfnamefont {M.~E. M. M. P. G. R.-T. F. S. G. Z.~F.}\
  \bibnamefont {Charbonneau}, \bibfnamefont {P.}},\ ed.,\ \href@noop {} {\emph
  {\bibinfo {title} {Spin Glass Theory and Far Beyond: Replica Symmetry
  Breaking After 40 Years}}}\ (\bibinfo  {publisher} {World Scientific},\
  \bibinfo {year} {2023})\BibitemShut {NoStop}%
\bibitem [{\citenamefont {Cannella}\ and\ \citenamefont
  {Mydosh}(1972)}]{cannella72}%
  \BibitemOpen
  \bibfield  {author} {\bibinfo {author} {\bibfnamefont {V.}~\bibnamefont
  {Cannella}}\ and\ \bibinfo {author} {\bibfnamefont {J.~A.}\ \bibnamefont
  {Mydosh}},\ }\bibfield  {title} {\bibinfo {title} {Magnetic ordering in
  gold-iron alloys},\ }\href {https://doi.org/10.1103/PhysRevB.6.4220}
  {\bibfield  {journal} {\bibinfo  {journal} {Phys. Rev. B}\ }\textbf {\bibinfo
  {volume} {6}},\ \bibinfo {pages} {4220} (\bibinfo {year} {1972})}\BibitemShut
  {NoStop}%
\bibitem [{\citenamefont {Lamelas}\ \emph {et~al.}(1995)\citenamefont
  {Lamelas}, \citenamefont {Werner}, \citenamefont {Shapiro},\ and\
  \citenamefont {Mydosh}}]{lamelas95}%
  \BibitemOpen
  \bibfield  {author} {\bibinfo {author} {\bibfnamefont {F.~J.}\ \bibnamefont
  {Lamelas}}, \bibinfo {author} {\bibfnamefont {S.~A.}\ \bibnamefont {Werner}},
  \bibinfo {author} {\bibfnamefont {S.~M.}\ \bibnamefont {Shapiro}},\ and\
  \bibinfo {author} {\bibfnamefont {J.~A.}\ \bibnamefont {Mydosh}},\ }\bibfield
   {title} {\bibinfo {title} {Intrinsic spin-density-wave magnetism in cu-mn
  alloys},\ }\href {https://doi.org/10.1103/PhysRevB.51.621} {\bibfield
  {journal} {\bibinfo  {journal} {Phys. Rev. B}\ }\textbf {\bibinfo {volume}
  {51}},\ \bibinfo {pages} {621} (\bibinfo {year} {1995})}\BibitemShut
  {NoStop}%
\bibitem [{\citenamefont {Ruderman}\ and\ \citenamefont
  {Kittel}(1954)}]{ruderman54}%
  \BibitemOpen
  \bibfield  {author} {\bibinfo {author} {\bibfnamefont {M.~A.}\ \bibnamefont
  {Ruderman}}\ and\ \bibinfo {author} {\bibfnamefont {C.}~\bibnamefont
  {Kittel}},\ }\bibfield  {title} {\bibinfo {title} {Indirect exchange coupling
  of nuclear magnetic moments by conduction electrons},\ }\href
  {https://doi.org/10.1103/PhysRev.96.99} {\bibfield  {journal} {\bibinfo
  {journal} {Phys. Rev.}\ }\textbf {\bibinfo {volume} {96}},\ \bibinfo {pages}
  {99} (\bibinfo {year} {1954})}\BibitemShut {NoStop}%
\bibitem [{\citenamefont {Kasuya}(1956)}]{kasuya56}%
  \BibitemOpen
  \bibfield  {author} {\bibinfo {author} {\bibfnamefont {T.}~\bibnamefont
  {Kasuya}},\ }\bibfield  {title} {\bibinfo {title} {{A Theory of Metallic
  Ferro- and Antiferromagnetism on Zener's Model}},\ }\href
  {https://doi.org/10.1143/PTP.16.45} {\bibfield  {journal} {\bibinfo
  {journal} {Progress of Theoretical Physics}\ }\textbf {\bibinfo {volume}
  {16}},\ \bibinfo {pages} {45} (\bibinfo {year} {1956})}\BibitemShut {NoStop}%
\bibitem [{\citenamefont {Yosida}(1957)}]{yosida57}%
  \BibitemOpen
  \bibfield  {author} {\bibinfo {author} {\bibfnamefont {K.}~\bibnamefont
  {Yosida}},\ }\bibfield  {title} {\bibinfo {title} {Magnetic properties of
  {C}u-{M}n alloys},\ }\href {https://doi.org/10.1103/PhysRev.106.893}
  {\bibfield  {journal} {\bibinfo  {journal} {Phys. Rev.}\ }\textbf {\bibinfo
  {volume} {106}},\ \bibinfo {pages} {893} (\bibinfo {year}
  {1957})}\BibitemShut {NoStop}%
\bibitem [{\citenamefont {Marx}\ and\ \citenamefont {Hutter}(2009)}]{marx09}%
  \BibitemOpen
  \bibfield  {author} {\bibinfo {author} {\bibfnamefont {D.}~\bibnamefont
  {Marx}}\ and\ \bibinfo {author} {\bibfnamefont {J.}~\bibnamefont {Hutter}},\
  }\href@noop {} {\emph {\bibinfo {title} {Ab initio molecular dynamics: basic
  theory and advanced methods}}}\ (\bibinfo  {publisher} {Cambridge University
  Press},\ \bibinfo {year} {2009})\BibitemShut {NoStop}%
\bibitem [{\citenamefont {Behler}\ and\ \citenamefont
  {Parrinello}(2007)}]{behler07}%
  \BibitemOpen
  \bibfield  {author} {\bibinfo {author} {\bibfnamefont {J.}~\bibnamefont
  {Behler}}\ and\ \bibinfo {author} {\bibfnamefont {M.}~\bibnamefont
  {Parrinello}},\ }\bibfield  {title} {\bibinfo {title} {Generalized
  neural-network representation of high-dimensional potential-energy
  surfaces},\ }\href {https://doi.org/10.1103/PhysRevLett.98.146401} {\bibfield
   {journal} {\bibinfo  {journal} {Phys. Rev. Lett.}\ }\textbf {\bibinfo
  {volume} {98}},\ \bibinfo {pages} {146401} (\bibinfo {year}
  {2007})}\BibitemShut {NoStop}%
\bibitem [{\citenamefont {Bart\'ok}\ \emph {et~al.}(2010)\citenamefont
  {Bart\'ok}, \citenamefont {Payne}, \citenamefont {Kondor},\ and\
  \citenamefont {Cs\'anyi}}]{bartok10}%
  \BibitemOpen
  \bibfield  {author} {\bibinfo {author} {\bibfnamefont {A.~P.}\ \bibnamefont
  {Bart\'ok}}, \bibinfo {author} {\bibfnamefont {M.~C.}\ \bibnamefont {Payne}},
  \bibinfo {author} {\bibfnamefont {R.}~\bibnamefont {Kondor}},\ and\ \bibinfo
  {author} {\bibfnamefont {G.}~\bibnamefont {Cs\'anyi}},\ }\bibfield  {title}
  {\bibinfo {title} {Gaussian approximation potentials: The accuracy of quantum
  mechanics, without the electrons},\ }\href
  {https://doi.org/10.1103/PhysRevLett.104.136403} {\bibfield  {journal}
  {\bibinfo  {journal} {Phys. Rev. Lett.}\ }\textbf {\bibinfo {volume} {104}},\
  \bibinfo {pages} {136403} (\bibinfo {year} {2010})}\BibitemShut {NoStop}%
\bibitem [{\citenamefont {Li}\ \emph {et~al.}(2015)\citenamefont {Li},
  \citenamefont {Kermode},\ and\ \citenamefont {De~Vita}}]{li15}%
  \BibitemOpen
  \bibfield  {author} {\bibinfo {author} {\bibfnamefont {Z.}~\bibnamefont
  {Li}}, \bibinfo {author} {\bibfnamefont {J.~R.}\ \bibnamefont {Kermode}},\
  and\ \bibinfo {author} {\bibfnamefont {A.}~\bibnamefont {De~Vita}},\
  }\bibfield  {title} {\bibinfo {title} {Molecular dynamics with on-the-fly
  machine learning of quantum-mechanical forces},\ }\href
  {https://doi.org/10.1103/PhysRevLett.114.096405} {\bibfield  {journal}
  {\bibinfo  {journal} {Phys. Rev. Lett.}\ }\textbf {\bibinfo {volume} {114}},\
  \bibinfo {pages} {096405} (\bibinfo {year} {2015})}\BibitemShut {NoStop}%
\bibitem [{\citenamefont {Shapeev}(2016)}]{shapeev16}%
  \BibitemOpen
  \bibfield  {author} {\bibinfo {author} {\bibfnamefont {A.~V.}\ \bibnamefont
  {Shapeev}},\ }\bibfield  {title} {\bibinfo {title} {Moment tensor potentials:
  A class of systematically improvable interatomic potentials},\ }\href
  {https://doi.org/10.1137/15M1054183} {\bibfield  {journal} {\bibinfo
  {journal} {Multiscale Modeling \& Simulation}\ }\textbf {\bibinfo {volume}
  {14}},\ \bibinfo {pages} {1153} (\bibinfo {year} {2016})}\BibitemShut
  {NoStop}%
\bibitem [{\citenamefont {Botu}\ \emph {et~al.}(2017)\citenamefont {Botu},
  \citenamefont {Batra}, \citenamefont {Chapman},\ and\ \citenamefont
  {Ramprasad}}]{botu17}%
  \BibitemOpen
  \bibfield  {author} {\bibinfo {author} {\bibfnamefont {V.}~\bibnamefont
  {Botu}}, \bibinfo {author} {\bibfnamefont {R.}~\bibnamefont {Batra}},
  \bibinfo {author} {\bibfnamefont {J.}~\bibnamefont {Chapman}},\ and\ \bibinfo
  {author} {\bibfnamefont {R.}~\bibnamefont {Ramprasad}},\ }\bibfield  {title}
  {\bibinfo {title} {Machine learning force fields: Construction, validation,
  and outlook},\ }\href {https://doi.org/10.1021/acs.jpcc.6b10908} {\bibfield
  {journal} {\bibinfo  {journal} {The Journal of Physical Chemistry C}\
  }\textbf {\bibinfo {volume} {121}},\ \bibinfo {pages} {511} (\bibinfo {year}
  {2017})}\BibitemShut {NoStop}%
\bibitem [{\citenamefont {Smith}\ \emph {et~al.}(2017)\citenamefont {Smith},
  \citenamefont {Isayev},\ and\ \citenamefont {Roitberg}}]{smith17}%
  \BibitemOpen
  \bibfield  {author} {\bibinfo {author} {\bibfnamefont {J.~S.}\ \bibnamefont
  {Smith}}, \bibinfo {author} {\bibfnamefont {O.}~\bibnamefont {Isayev}},\ and\
  \bibinfo {author} {\bibfnamefont {A.~E.}\ \bibnamefont {Roitberg}},\
  }\bibfield  {title} {\bibinfo {title} {Ani-1: an extensible neural network
  potential with dft accuracy at force field computational cost},\ }\href
  {https://doi.org/10.1039/C6SC05720A} {\bibfield  {journal} {\bibinfo
  {journal} {Chem. Sci.}\ }\textbf {\bibinfo {volume} {8}},\ \bibinfo {pages}
  {3192} (\bibinfo {year} {2017})}\BibitemShut {NoStop}%
\bibitem [{\citenamefont {Zhang}\ \emph {et~al.}(2018)\citenamefont {Zhang},
  \citenamefont {Han}, \citenamefont {Wang}, \citenamefont {Car},\ and\
  \citenamefont {E}}]{zhang18}%
  \BibitemOpen
  \bibfield  {author} {\bibinfo {author} {\bibfnamefont {L.}~\bibnamefont
  {Zhang}}, \bibinfo {author} {\bibfnamefont {J.}~\bibnamefont {Han}}, \bibinfo
  {author} {\bibfnamefont {H.}~\bibnamefont {Wang}}, \bibinfo {author}
  {\bibfnamefont {R.}~\bibnamefont {Car}},\ and\ \bibinfo {author}
  {\bibfnamefont {W.}~\bibnamefont {E}},\ }\bibfield  {title} {\bibinfo {title}
  {Deep potential molecular dynamics: A scalable model with the accuracy of
  quantum mechanics},\ }\href {https://doi.org/10.1103/PhysRevLett.120.143001}
  {\bibfield  {journal} {\bibinfo  {journal} {Phys. Rev. Lett.}\ }\textbf
  {\bibinfo {volume} {120}},\ \bibinfo {pages} {143001} (\bibinfo {year}
  {2018})}\BibitemShut {NoStop}%
\bibitem [{\citenamefont {Behler}(2016)}]{behler16}%
  \BibitemOpen
  \bibfield  {author} {\bibinfo {author} {\bibfnamefont {J.}~\bibnamefont
  {Behler}},\ }\bibfield  {title} {\bibinfo {title} {{Perspective: Machine
  learning potentials for atomistic simulations}},\ }\href
  {https://doi.org/10.1063/1.4966192} {\bibfield  {journal} {\bibinfo
  {journal} {The Journal of Chemical Physics}\ }\textbf {\bibinfo {volume}
  {145}},\ \bibinfo {pages} {170901} (\bibinfo {year} {2016})}\BibitemShut
  {NoStop}%
\bibitem [{\citenamefont {Deringer}(2019)}]{deringer19}%
  \BibitemOpen
  \bibfield  {author} {\bibinfo {author} {\bibfnamefont {C.~M. A. C.~G.}\
  \bibnamefont {Deringer}, \bibfnamefont {V.~L.}},\ }\bibfield  {title}
  {\bibinfo {title} {Machine learning interatomic potentials as emerging tools
  for materials science},\ }\href
  {https://doi.org/https://doi.org/10.1002/adma.201902765} {\bibfield
  {journal} {\bibinfo  {journal} {Advanced Materials}\ }\textbf {\bibinfo
  {volume} {31}},\ \bibinfo {pages} {1902765} (\bibinfo {year}
  {2019})}\BibitemShut {NoStop}%
\bibitem [{\citenamefont {McGibbon}(2017)}]{mcgibbon17}%
  \BibitemOpen
  \bibfield  {author} {\bibinfo {author} {\bibfnamefont {T.~A. G. D. A. G.-S.
  K. H. F. H. C. L. K.-H. K. J. L. S. D.~E.}\ \bibnamefont {McGibbon},
  \bibfnamefont {R.~T.}},\ }\bibfield  {title} {\bibinfo {title} {Improving the
  accuracy of m\"oller-plesset perturbation theory with neural networks},\
  }\href {https://doi.org/10.1063/1.4986081} {\bibfield  {journal} {\bibinfo
  {journal} {The Journal of Chemical Physics}\ }\textbf {\bibinfo {volume}
  {147}},\ \bibinfo {pages} {161725} (\bibinfo {year} {2017})}\BibitemShut
  {NoStop}%
\bibitem [{\citenamefont {Suwa}\ \emph {et~al.}(2019)\citenamefont {Suwa},
  \citenamefont {Smith}, \citenamefont {Lubbers}, \citenamefont {Batista},
  \citenamefont {Chern},\ and\ \citenamefont {Barros}}]{suwa19}%
  \BibitemOpen
  \bibfield  {author} {\bibinfo {author} {\bibfnamefont {H.}~\bibnamefont
  {Suwa}}, \bibinfo {author} {\bibfnamefont {J.~S.}\ \bibnamefont {Smith}},
  \bibinfo {author} {\bibfnamefont {N.}~\bibnamefont {Lubbers}}, \bibinfo
  {author} {\bibfnamefont {C.~D.}\ \bibnamefont {Batista}}, \bibinfo {author}
  {\bibfnamefont {G.-W.}\ \bibnamefont {Chern}},\ and\ \bibinfo {author}
  {\bibfnamefont {K.}~\bibnamefont {Barros}},\ }\bibfield  {title} {\bibinfo
  {title} {Machine learning for molecular dynamics with strongly correlated
  electrons},\ }\href {https://doi.org/10.1103/PhysRevB.99.161107} {\bibfield
  {journal} {\bibinfo  {journal} {Phys. Rev. B}\ }\textbf {\bibinfo {volume}
  {99}},\ \bibinfo {pages} {161107} (\bibinfo {year} {2019})}\BibitemShut
  {NoStop}%
\bibitem [{\citenamefont {Chmiela}\ \emph {et~al.}(2017)\citenamefont
  {Chmiela}, \citenamefont {Tkatchenko}, \citenamefont {Sauceda}, \citenamefont
  {Poltavsky}, \citenamefont {Schütt},\ and\ \citenamefont
  {Müller}}]{chmiela17}%
  \BibitemOpen
  \bibfield  {author} {\bibinfo {author} {\bibfnamefont {S.}~\bibnamefont
  {Chmiela}}, \bibinfo {author} {\bibfnamefont {A.}~\bibnamefont {Tkatchenko}},
  \bibinfo {author} {\bibfnamefont {H.~E.}\ \bibnamefont {Sauceda}}, \bibinfo
  {author} {\bibfnamefont {I.}~\bibnamefont {Poltavsky}}, \bibinfo {author}
  {\bibfnamefont {K.~T.}\ \bibnamefont {Schütt}},\ and\ \bibinfo {author}
  {\bibfnamefont {K.-R.}\ \bibnamefont {Müller}},\ }\bibfield  {title}
  {\bibinfo {title} {Machine learning of accurate energy-conserving molecular
  force fields},\ }\href {https://doi.org/10.1126/sciadv.1603015} {\bibfield
  {journal} {\bibinfo  {journal} {Science Advances}\ }\textbf {\bibinfo
  {volume} {3}},\ \bibinfo {pages} {e1603015} (\bibinfo {year}
  {2017})}\BibitemShut {NoStop}%
\bibitem [{\citenamefont {Chmiela}\ \emph {et~al.}(2018)\citenamefont
  {Chmiela}, \citenamefont {Sauceda}, \citenamefont {M{\"u}ller},\ and\
  \citenamefont {Tkatchenko}}]{chmiela18}%
  \BibitemOpen
  \bibfield  {author} {\bibinfo {author} {\bibfnamefont {S.}~\bibnamefont
  {Chmiela}}, \bibinfo {author} {\bibfnamefont {H.~E.}\ \bibnamefont
  {Sauceda}}, \bibinfo {author} {\bibfnamefont {K.-R.}\ \bibnamefont
  {M{\"u}ller}},\ and\ \bibinfo {author} {\bibfnamefont {A.}~\bibnamefont
  {Tkatchenko}},\ }\bibfield  {title} {\bibinfo {title} {Towards exact
  molecular dynamics simulations with machine-learned force fields},\ }\href
  {https://doi.org/10.1038/s41467-018-06169-2} {\bibfield  {journal} {\bibinfo
  {journal} {Nature Communications}\ }\textbf {\bibinfo {volume} {9}},\
  \bibinfo {pages} {3887} (\bibinfo {year} {2018})}\BibitemShut {NoStop}%
\bibitem [{\citenamefont {Sauceda}\ \emph {et~al.}(2020)\citenamefont
  {Sauceda}, \citenamefont {Gastegger}, \citenamefont {Chmiela}, \citenamefont
  {Müller},\ and\ \citenamefont {Tkatchenko}}]{sauceda20}%
  \BibitemOpen
  \bibfield  {author} {\bibinfo {author} {\bibfnamefont {H.~E.}\ \bibnamefont
  {Sauceda}}, \bibinfo {author} {\bibfnamefont {M.}~\bibnamefont {Gastegger}},
  \bibinfo {author} {\bibfnamefont {S.}~\bibnamefont {Chmiela}}, \bibinfo
  {author} {\bibfnamefont {K.-R.}\ \bibnamefont {Müller}},\ and\ \bibinfo
  {author} {\bibfnamefont {A.}~\bibnamefont {Tkatchenko}},\ }\bibfield  {title}
  {\bibinfo {title} {{Molecular force fields with gradient-domain machine
  learning (GDML): Comparison and synergies with classical force fields}},\
  }\href {https://doi.org/10.1063/5.0023005} {\bibfield  {journal} {\bibinfo
  {journal} {The Journal of Chemical Physics}\ }\textbf {\bibinfo {volume}
  {153}},\ \bibinfo {pages} {124109} (\bibinfo {year} {2020})}\BibitemShut
  {NoStop}%
\bibitem [{\citenamefont {Zhang}\ \emph {et~al.}(2020)\citenamefont {Zhang},
  \citenamefont {Saha},\ and\ \citenamefont {Chern}}]{zhang20}%
  \BibitemOpen
  \bibfield  {author} {\bibinfo {author} {\bibfnamefont {P.}~\bibnamefont
  {Zhang}}, \bibinfo {author} {\bibfnamefont {P.}~\bibnamefont {Saha}},\ and\
  \bibinfo {author} {\bibfnamefont {G.-W.}\ \bibnamefont {Chern}},\ }\href@noop
  {} {\bibinfo {title} {Machine learning dynamics of phase separation in
  correlated electron magnets}} (\bibinfo {year} {2020}),\ \Eprint
  {https://arxiv.org/abs/2006.04205} {arXiv:2006.04205 [cond-mat.str-el]}
  \BibitemShut {NoStop}%
\bibitem [{\citenamefont {Zhang}\ and\ \citenamefont {Chern}(2021)}]{zhang21}%
  \BibitemOpen
  \bibfield  {author} {\bibinfo {author} {\bibfnamefont {P.}~\bibnamefont
  {Zhang}}\ and\ \bibinfo {author} {\bibfnamefont {G.-W.}\ \bibnamefont
  {Chern}},\ }\bibfield  {title} {\bibinfo {title} {Arrested phase separation
  in double-exchange models: Large-scale simulation enabled by machine
  learning},\ }\href {https://doi.org/10.1103/PhysRevLett.127.146401}
  {\bibfield  {journal} {\bibinfo  {journal} {Phys. Rev. Lett.}\ }\textbf
  {\bibinfo {volume} {127}},\ \bibinfo {pages} {146401} (\bibinfo {year}
  {2021})}\BibitemShut {NoStop}%
\bibitem [{\citenamefont {Zhang}\ \emph
  {et~al.}(2022{\natexlab{a}})\citenamefont {Zhang}, \citenamefont {Zhang},\
  and\ \citenamefont {Chern}}]{zhang22}%
  \BibitemOpen
  \bibfield  {author} {\bibinfo {author} {\bibfnamefont {P.}~\bibnamefont
  {Zhang}}, \bibinfo {author} {\bibfnamefont {S.}~\bibnamefont {Zhang}},\ and\
  \bibinfo {author} {\bibfnamefont {G.-W.}\ \bibnamefont {Chern}},\ }\href@noop
  {} {\bibinfo {title} {Descriptors for machine learning model of generalized
  force field in condensed matter systems}} (\bibinfo {year}
  {2022}{\natexlab{a}}),\ \Eprint {https://arxiv.org/abs/2201.00798}
  {arXiv:2201.00798 [cond-mat.str-el]} \BibitemShut {NoStop}%
\bibitem [{\citenamefont {Zhang}\ \emph
  {et~al.}(2022{\natexlab{b}})\citenamefont {Zhang}, \citenamefont {Zhang},\
  and\ \citenamefont {Chern}}]{zhang22b}%
  \BibitemOpen
  \bibfield  {author} {\bibinfo {author} {\bibfnamefont {S.}~\bibnamefont
  {Zhang}}, \bibinfo {author} {\bibfnamefont {P.}~\bibnamefont {Zhang}},\ and\
  \bibinfo {author} {\bibfnamefont {G.-W.}\ \bibnamefont {Chern}},\ }\bibfield
  {title} {\bibinfo {title} {Anomalous phase separation in a correlated
  electron system: Machine-learning enabled large-scale kinetic monte carlo
  simulations},\ }\href {https://doi.org/10.1073/pnas.2119957119} {\bibfield
  {journal} {\bibinfo  {journal} {Proceedings of the National Academy of
  Sciences}\ }\textbf {\bibinfo {volume} {119}},\ \bibinfo {pages}
  {e2119957119} (\bibinfo {year} {2022}{\natexlab{b}})}\BibitemShut {NoStop}%
\bibitem [{\citenamefont {Zhang}\ and\ \citenamefont {Chern}(2023)}]{zhang23}%
  \BibitemOpen
  \bibfield  {author} {\bibinfo {author} {\bibfnamefont {P.}~\bibnamefont
  {Zhang}}\ and\ \bibinfo {author} {\bibfnamefont {G.-W.}\ \bibnamefont
  {Chern}},\ }\bibfield  {title} {\bibinfo {title} {Machine learning
  nonequilibrium electron forces for spin dynamics of itinerant magnets},\
  }\href {https://doi.org/10.1038/s41524-023-00990-0} {\bibfield  {journal}
  {\bibinfo  {journal} {npj Computational Materials}\ }\textbf {\bibinfo
  {volume} {9}},\ \bibinfo {pages} {32} (\bibinfo {year} {2023})}\BibitemShut
  {NoStop}%
\bibitem [{\citenamefont {Cheng}\ \emph
  {et~al.}(2023{\natexlab{a}})\citenamefont {Cheng}, \citenamefont {Zhang},\
  and\ \citenamefont {Chern}}]{cheng23}%
  \BibitemOpen
  \bibfield  {author} {\bibinfo {author} {\bibfnamefont {C.}~\bibnamefont
  {Cheng}}, \bibinfo {author} {\bibfnamefont {S.}~\bibnamefont {Zhang}},\ and\
  \bibinfo {author} {\bibfnamefont {G.-W.}\ \bibnamefont {Chern}},\ }\bibfield
  {title} {\bibinfo {title} {Machine learning for phase ordering dynamics of
  charge density waves},\ }\href {https://doi.org/10.1103/PhysRevB.108.014301}
  {\bibfield  {journal} {\bibinfo  {journal} {Phys. Rev. B}\ }\textbf {\bibinfo
  {volume} {108}},\ \bibinfo {pages} {014301} (\bibinfo {year}
  {2023}{\natexlab{a}})}\BibitemShut {NoStop}%
\bibitem [{\citenamefont {Cheng}\ \emph
  {et~al.}(2023{\natexlab{b}})\citenamefont {Cheng}, \citenamefont {Zhang},
  \citenamefont {Nguyen}, \citenamefont {Azarfar}, \citenamefont {Chern},\ and\
  \citenamefont {Baek}}]{cheng23b}%
  \BibitemOpen
  \bibfield  {author} {\bibinfo {author} {\bibfnamefont {X.}~\bibnamefont
  {Cheng}}, \bibinfo {author} {\bibfnamefont {S.}~\bibnamefont {Zhang}},
  \bibinfo {author} {\bibfnamefont {P.~C.~H.}\ \bibnamefont {Nguyen}}, \bibinfo
  {author} {\bibfnamefont {S.}~\bibnamefont {Azarfar}}, \bibinfo {author}
  {\bibfnamefont {G.-W.}\ \bibnamefont {Chern}},\ and\ \bibinfo {author}
  {\bibfnamefont {S.~S.}\ \bibnamefont {Baek}},\ }\bibfield  {title} {\bibinfo
  {title} {Convolutional neural networks for large-scale dynamical modeling of
  itinerant magnets},\ }\href
  {https://doi.org/10.1103/PhysRevResearch.5.033188} {\bibfield  {journal}
  {\bibinfo  {journal} {Phys. Rev. Res.}\ }\textbf {\bibinfo {volume} {5}},\
  \bibinfo {pages} {033188} (\bibinfo {year} {2023}{\natexlab{b}})}\BibitemShut
  {NoStop}%
\bibitem [{\citenamefont {Kohn}(1996)}]{kohn96}%
  \BibitemOpen
  \bibfield  {author} {\bibinfo {author} {\bibfnamefont {W.}~\bibnamefont
  {Kohn}},\ }\bibfield  {title} {\bibinfo {title} {Density functional and
  density matrix method scaling linearly with the number of atoms},\ }\href
  {https://doi.org/10.1103/PhysRevLett.76.3168} {\bibfield  {journal} {\bibinfo
   {journal} {Phys. Rev. Lett.}\ }\textbf {\bibinfo {volume} {76}},\ \bibinfo
  {pages} {3168} (\bibinfo {year} {1996})}\BibitemShut {NoStop}%
\bibitem [{\citenamefont {Prodan}\ and\ \citenamefont {Kohn}(2005)}]{prodan05}%
  \BibitemOpen
  \bibfield  {author} {\bibinfo {author} {\bibfnamefont {E.}~\bibnamefont
  {Prodan}}\ and\ \bibinfo {author} {\bibfnamefont {W.}~\bibnamefont {Kohn}},\
  }\bibfield  {title} {\bibinfo {title} {Nearsightedness of electronic
  matter},\ }\href {https://doi.org/10.1073/pnas.0505436102} {\bibfield
  {journal} {\bibinfo  {journal} {Proceedings of the National Academy of
  Sciences}\ }\textbf {\bibinfo {volume} {102}},\ \bibinfo {pages} {11635}
  (\bibinfo {year} {2005})}\BibitemShut {NoStop}%
\bibitem [{\citenamefont {Behler}(2011)}]{behler11}%
  \BibitemOpen
  \bibfield  {author} {\bibinfo {author} {\bibfnamefont {J.}~\bibnamefont
  {Behler}},\ }\bibfield  {title} {\bibinfo {title} {{Atom-centered symmetry
  functions for constructing high-dimensional neural network potentials}},\
  }\href {https://doi.org/10.1063/1.3553717} {\bibfield  {journal} {\bibinfo
  {journal} {The Journal of Chemical Physics}\ }\textbf {\bibinfo {volume}
  {134}},\ \bibinfo {pages} {074106} (\bibinfo {year} {2011})}\BibitemShut
  {NoStop}%
\bibitem [{\citenamefont {Ghiringhelli}\ \emph {et~al.}(2015)\citenamefont
  {Ghiringhelli}, \citenamefont {Vybiral}, \citenamefont {Levchenko},
  \citenamefont {Draxl},\ and\ \citenamefont {Scheffler}}]{ghiringhelli15}%
  \BibitemOpen
  \bibfield  {author} {\bibinfo {author} {\bibfnamefont {L.~M.}\ \bibnamefont
  {Ghiringhelli}}, \bibinfo {author} {\bibfnamefont {J.}~\bibnamefont
  {Vybiral}}, \bibinfo {author} {\bibfnamefont {S.~V.}\ \bibnamefont
  {Levchenko}}, \bibinfo {author} {\bibfnamefont {C.}~\bibnamefont {Draxl}},\
  and\ \bibinfo {author} {\bibfnamefont {M.}~\bibnamefont {Scheffler}},\
  }\bibfield  {title} {\bibinfo {title} {Big data of materials science:
  Critical role of the descriptor},\ }\href
  {https://doi.org/10.1103/PhysRevLett.114.105503} {\bibfield  {journal}
  {\bibinfo  {journal} {Phys. Rev. Lett.}\ }\textbf {\bibinfo {volume} {114}},\
  \bibinfo {pages} {105503} (\bibinfo {year} {2015})}\BibitemShut {NoStop}%
\bibitem [{\citenamefont {Bart\'ok}\ \emph {et~al.}(2013)\citenamefont
  {Bart\'ok}, \citenamefont {Kondor},\ and\ \citenamefont
  {Cs\'anyi}}]{bartok13}%
  \BibitemOpen
  \bibfield  {author} {\bibinfo {author} {\bibfnamefont {A.~P.}\ \bibnamefont
  {Bart\'ok}}, \bibinfo {author} {\bibfnamefont {R.}~\bibnamefont {Kondor}},\
  and\ \bibinfo {author} {\bibfnamefont {G.}~\bibnamefont {Cs\'anyi}},\
  }\bibfield  {title} {\bibinfo {title} {On representing chemical
  environments},\ }\href {https://doi.org/10.1103/PhysRevB.87.184115}
  {\bibfield  {journal} {\bibinfo  {journal} {Phys. Rev. B}\ }\textbf {\bibinfo
  {volume} {87}},\ \bibinfo {pages} {184115} (\bibinfo {year}
  {2013})}\BibitemShut {NoStop}%
\bibitem [{\citenamefont {Drautz}(2019)}]{drautz19}%
  \BibitemOpen
  \bibfield  {author} {\bibinfo {author} {\bibfnamefont {R.}~\bibnamefont
  {Drautz}},\ }\bibfield  {title} {\bibinfo {title} {Atomic cluster expansion
  for accurate and transferable interatomic potentials},\ }\href
  {https://doi.org/10.1103/PhysRevB.99.014104} {\bibfield  {journal} {\bibinfo
  {journal} {Phys. Rev. B}\ }\textbf {\bibinfo {volume} {99}},\ \bibinfo
  {pages} {014104} (\bibinfo {year} {2019})}\BibitemShut {NoStop}%
\bibitem [{\citenamefont {Himanen}\ \emph {et~al.}(2020)\citenamefont
  {Himanen}, \citenamefont {Jäger}, \citenamefont {Morooka}, \citenamefont
  {{Federici Canova}}, \citenamefont {Ranawat}, \citenamefont {Gao},
  \citenamefont {Rinke},\ and\ \citenamefont {Foster}}]{himanen20}%
  \BibitemOpen
  \bibfield  {author} {\bibinfo {author} {\bibfnamefont {L.}~\bibnamefont
  {Himanen}}, \bibinfo {author} {\bibfnamefont {M.~O.}\ \bibnamefont {Jäger}},
  \bibinfo {author} {\bibfnamefont {E.~V.}\ \bibnamefont {Morooka}}, \bibinfo
  {author} {\bibfnamefont {F.}~\bibnamefont {{Federici Canova}}}, \bibinfo
  {author} {\bibfnamefont {Y.~S.}\ \bibnamefont {Ranawat}}, \bibinfo {author}
  {\bibfnamefont {D.~Z.}\ \bibnamefont {Gao}}, \bibinfo {author} {\bibfnamefont
  {P.}~\bibnamefont {Rinke}},\ and\ \bibinfo {author} {\bibfnamefont {A.~S.}\
  \bibnamefont {Foster}},\ }\bibfield  {title} {\bibinfo {title} {Dscribe:
  Library of descriptors for machine learning in materials science},\ }\href
  {https://doi.org/https://doi.org/10.1016/j.cpc.2019.106949} {\bibfield
  {journal} {\bibinfo  {journal} {Computer Physics Communications}\ }\textbf
  {\bibinfo {volume} {247}},\ \bibinfo {pages} {106949} (\bibinfo {year}
  {2020})}\BibitemShut {NoStop}%
\bibitem [{\citenamefont {Huo}\ and\ \citenamefont {Rupp}(2022)}]{huo22}%
  \BibitemOpen
  \bibfield  {author} {\bibinfo {author} {\bibfnamefont {H.}~\bibnamefont
  {Huo}}\ and\ \bibinfo {author} {\bibfnamefont {M.}~\bibnamefont {Rupp}},\
  }\bibfield  {title} {\bibinfo {title} {Unified representation of molecules
  and crystals for machine learning},\ }\href
  {https://doi.org/10.1088/2632-2153/aca005} {\bibfield  {journal} {\bibinfo
  {journal} {Machine Learning: Science and Technology}\ }\textbf {\bibinfo
  {volume} {3}},\ \bibinfo {pages} {045017} (\bibinfo {year}
  {2022})}\BibitemShut {NoStop}%
\bibitem [{\citenamefont {Skubic}\ \emph {et~al.}(2009)\citenamefont {Skubic},
  \citenamefont {Peil}, \citenamefont {Hellsvik}, \citenamefont {Nordblad},
  \citenamefont {Nordstr\"om},\ and\ \citenamefont {Eriksson}}]{skubic09}%
  \BibitemOpen
  \bibfield  {author} {\bibinfo {author} {\bibfnamefont {B.}~\bibnamefont
  {Skubic}}, \bibinfo {author} {\bibfnamefont {O.~E.}\ \bibnamefont {Peil}},
  \bibinfo {author} {\bibfnamefont {J.}~\bibnamefont {Hellsvik}}, \bibinfo
  {author} {\bibfnamefont {P.}~\bibnamefont {Nordblad}}, \bibinfo {author}
  {\bibfnamefont {L.}~\bibnamefont {Nordstr\"om}},\ and\ \bibinfo {author}
  {\bibfnamefont {O.}~\bibnamefont {Eriksson}},\ }\bibfield  {title} {\bibinfo
  {title} {Atomistic spin dynamics of the cu-mn spin-glass alloy},\ }\href
  {https://doi.org/10.1103/PhysRevB.79.024411} {\bibfield  {journal} {\bibinfo
  {journal} {Phys. Rev. B}\ }\textbf {\bibinfo {volume} {79}},\ \bibinfo
  {pages} {024411} (\bibinfo {year} {2009})}\BibitemShut {NoStop}%
\bibitem [{\citenamefont {Hellsvik}(2010)}]{hellsvik10}%
  \BibitemOpen
  \bibfield  {author} {\bibinfo {author} {\bibfnamefont {J.}~\bibnamefont
  {Hellsvik}},\ }\bibfield  {title} {\bibinfo {title} {Atomistic spin dynamics
  simulations on mn-doped gaas and cumn},\ }\href
  {https://doi.org/10.1088/1742-6596/200/7/072040} {\bibfield  {journal}
  {\bibinfo  {journal} {Journal of Physics: Conference Series}\ }\textbf
  {\bibinfo {volume} {200}},\ \bibinfo {pages} {072040} (\bibinfo {year}
  {2010})}\BibitemShut {NoStop}%
\bibitem [{\citenamefont {Peil}\ \emph {et~al.}(2008)\citenamefont {Peil},
  \citenamefont {Ruban},\ and\ \citenamefont {Johansson}}]{peil08}%
  \BibitemOpen
  \bibfield  {author} {\bibinfo {author} {\bibfnamefont {O.~E.}\ \bibnamefont
  {Peil}}, \bibinfo {author} {\bibfnamefont {A.~V.}\ \bibnamefont {Ruban}},\
  and\ \bibinfo {author} {\bibfnamefont {B.}~\bibnamefont {Johansson}},\
  }\bibfield  {title} {\bibinfo {title} {Detailed ab initio calculations of the
  structure and magnetic state of a metallic spin glass},\ }\href
  {https://doi.org/10.1088/1367-2630/10/8/083026} {\bibfield  {journal}
  {\bibinfo  {journal} {New Journal of Physics}\ }\textbf {\bibinfo {volume}
  {10}},\ \bibinfo {pages} {083026} (\bibinfo {year} {2008})}\BibitemShut
  {NoStop}%
\bibitem [{\citenamefont {Ling}\ \emph {et~al.}(1994)\citenamefont {Ling},
  \citenamefont {Staunton},\ and\ \citenamefont {Johnson}}]{ling94}%
  \BibitemOpen
  \bibfield  {author} {\bibinfo {author} {\bibfnamefont {M.~F.}\ \bibnamefont
  {Ling}}, \bibinfo {author} {\bibfnamefont {J.~B.}\ \bibnamefont {Staunton}},\
  and\ \bibinfo {author} {\bibfnamefont {D.~D.}\ \bibnamefont {Johnson}},\
  }\bibfield  {title} {\bibinfo {title} {A 'first-principles' theory for
  magnetic correlations and atomic short-range order in paramagnetic alloys.
  ii. application to cumn},\ }\href
  {https://doi.org/10.1088/0953-8984/6/30/017} {\bibfield  {journal} {\bibinfo
  {journal} {Journal of Physics: Condensed Matter}\ }\textbf {\bibinfo {volume}
  {6}},\ \bibinfo {pages} {6001} (\bibinfo {year} {1994})}\BibitemShut
  {NoStop}%
\bibitem [{\citenamefont {Ruban}\ \emph {et~al.}(2004)\citenamefont {Ruban},
  \citenamefont {Shallcross}, \citenamefont {Simak},\ and\ \citenamefont
  {Skriver}}]{ruban04}%
  \BibitemOpen
  \bibfield  {author} {\bibinfo {author} {\bibfnamefont {A.~V.}\ \bibnamefont
  {Ruban}}, \bibinfo {author} {\bibfnamefont {S.}~\bibnamefont {Shallcross}},
  \bibinfo {author} {\bibfnamefont {S.~I.}\ \bibnamefont {Simak}},\ and\
  \bibinfo {author} {\bibfnamefont {H.~L.}\ \bibnamefont {Skriver}},\
  }\bibfield  {title} {\bibinfo {title} {Atomic and magnetic configurational
  energetics by the generalized perturbation method},\ }\href
  {https://doi.org/10.1103/PhysRevB.70.125115} {\bibfield  {journal} {\bibinfo
  {journal} {Phys. Rev. B}\ }\textbf {\bibinfo {volume} {70}},\ \bibinfo
  {pages} {125115} (\bibinfo {year} {2004})}\BibitemShut {NoStop}%
\bibitem [{\citenamefont {Anderson}(1961)}]{anderson61}%
  \BibitemOpen
  \bibfield  {author} {\bibinfo {author} {\bibfnamefont {P.~W.}\ \bibnamefont
  {Anderson}},\ }\bibfield  {title} {\bibinfo {title} {Localized magnetic
  states in metals},\ }\href {https://doi.org/10.1103/PhysRev.124.41}
  {\bibfield  {journal} {\bibinfo  {journal} {Phys. Rev.}\ }\textbf {\bibinfo
  {volume} {124}},\ \bibinfo {pages} {41} (\bibinfo {year} {1961})}\BibitemShut
  {NoStop}%
\bibitem [{\citenamefont {Kondor}(2007)}]{kondor07}%
  \BibitemOpen
  \bibfield  {author} {\bibinfo {author} {\bibfnamefont {R.}~\bibnamefont
  {Kondor}},\ }\href@noop {} {\bibinfo {title} {A novel set of rotationally and
  translationally invariant features for images based on the non-commutative
  bispectrum}} (\bibinfo {year} {2007}),\ \Eprint
  {https://arxiv.org/abs/cs/0701127} {arXiv:cs/0701127 [cs.CV]} \BibitemShut
  {NoStop}%
\bibitem [{\citenamefont {Drautz}(2020)}]{drautz20}%
  \BibitemOpen
  \bibfield  {author} {\bibinfo {author} {\bibfnamefont {R.}~\bibnamefont
  {Drautz}},\ }\bibfield  {title} {\bibinfo {title} {Atomic cluster expansion
  of scalar, vectorial, and tensorial properties including magnetism and charge
  transfer},\ }\href {https://doi.org/10.1103/PhysRevB.102.024104} {\bibfield
  {journal} {\bibinfo  {journal} {Phys. Rev. B}\ }\textbf {\bibinfo {volume}
  {102}},\ \bibinfo {pages} {024104} (\bibinfo {year} {2020})}\BibitemShut
  {NoStop}%
\bibitem [{\citenamefont {Eckhoff}\ and\ \citenamefont
  {Behler}(2021)}]{eckhoff21}%
  \BibitemOpen
  \bibfield  {author} {\bibinfo {author} {\bibfnamefont {M.}~\bibnamefont
  {Eckhoff}}\ and\ \bibinfo {author} {\bibfnamefont {J.}~\bibnamefont
  {Behler}},\ }\bibfield  {title} {\bibinfo {title} {High-dimensional neural
  network potentials for magnetic systems using spin-dependent atom-centered
  symmetry functions},\ }\href {https://doi.org/10.1038/s41524-021-00636-z}
  {\bibfield  {journal} {\bibinfo  {journal} {npj Computational Materials}\
  }\textbf {\bibinfo {volume} {7}},\ \bibinfo {pages} {170} (\bibinfo {year}
  {2021})}\BibitemShut {NoStop}%
\bibitem [{\citenamefont {Arale~Br\"annvall}\ \emph {et~al.}(2022)\citenamefont
  {Arale~Br\"annvall}, \citenamefont {Gambino}, \citenamefont {Armiento},\ and\
  \citenamefont {Alling}}]{brannvall22}%
  \BibitemOpen
  \bibfield  {author} {\bibinfo {author} {\bibfnamefont {M.}~\bibnamefont
  {Arale~Br\"annvall}}, \bibinfo {author} {\bibfnamefont {D.}~\bibnamefont
  {Gambino}}, \bibinfo {author} {\bibfnamefont {R.}~\bibnamefont {Armiento}},\
  and\ \bibinfo {author} {\bibfnamefont {B.}~\bibnamefont {Alling}},\
  }\bibfield  {title} {\bibinfo {title} {Machine learning approach for
  longitudinal spin fluctuation effects in bcc fe at ${T}_{c}$ and under
  earth-core conditions},\ }\href {https://doi.org/10.1103/PhysRevB.105.144417}
  {\bibfield  {journal} {\bibinfo  {journal} {Phys. Rev. B}\ }\textbf {\bibinfo
  {volume} {105}},\ \bibinfo {pages} {144417} (\bibinfo {year}
  {2022})}\BibitemShut {NoStop}%
\bibitem [{\citenamefont {Domina}\ \emph {et~al.}(2022)\citenamefont {Domina},
  \citenamefont {Cobelli},\ and\ \citenamefont {Sanvito}}]{domina22}%
  \BibitemOpen
  \bibfield  {author} {\bibinfo {author} {\bibfnamefont {M.}~\bibnamefont
  {Domina}}, \bibinfo {author} {\bibfnamefont {M.}~\bibnamefont {Cobelli}},\
  and\ \bibinfo {author} {\bibfnamefont {S.}~\bibnamefont {Sanvito}},\
  }\bibfield  {title} {\bibinfo {title} {Spectral neighbor representation for
  vector fields: Machine learning potentials including spin},\ }\href
  {https://doi.org/10.1103/PhysRevB.105.214439} {\bibfield  {journal} {\bibinfo
   {journal} {Phys. Rev. B}\ }\textbf {\bibinfo {volume} {105}},\ \bibinfo
  {pages} {214439} (\bibinfo {year} {2022})}\BibitemShut {NoStop}%
\bibitem [{\citenamefont {Novikov}\ \emph {et~al.}(2022)\citenamefont
  {Novikov}, \citenamefont {Grabowski}, \citenamefont {K{\"o}rmann},\ and\
  \citenamefont {Shapeev}}]{novikov22}%
  \BibitemOpen
  \bibfield  {author} {\bibinfo {author} {\bibfnamefont {I.}~\bibnamefont
  {Novikov}}, \bibinfo {author} {\bibfnamefont {B.}~\bibnamefont {Grabowski}},
  \bibinfo {author} {\bibfnamefont {F.}~\bibnamefont {K{\"o}rmann}},\ and\
  \bibinfo {author} {\bibfnamefont {A.}~\bibnamefont {Shapeev}},\ }\bibfield
  {title} {\bibinfo {title} {Magnetic moment tensor potentials for collinear
  spin-polarized materials reproduce different magnetic states of bcc fe},\
  }\href {https://doi.org/10.1038/s41524-022-00696-9} {\bibfield  {journal}
  {\bibinfo  {journal} {npj Computational Materials}\ }\textbf {\bibinfo
  {volume} {8}},\ \bibinfo {pages} {13} (\bibinfo {year} {2022})}\BibitemShut
  {NoStop}%
\bibitem [{\citenamefont {Chapman}\ and\ \citenamefont {Ma}(2022)}]{chapman22}%
  \BibitemOpen
  \bibfield  {author} {\bibinfo {author} {\bibfnamefont {J.~B.~J.}\
  \bibnamefont {Chapman}}\ and\ \bibinfo {author} {\bibfnamefont {P.-W.}\
  \bibnamefont {Ma}},\ }\bibfield  {title} {\bibinfo {title} {A machine-learned
  spin-lattice potential for dynamic simulations of defective magnetic iron},\
  }\href {https://doi.org/10.1038/s41598-022-25682-5} {\bibfield  {journal}
  {\bibinfo  {journal} {Scientific Reports}\ }\textbf {\bibinfo {volume}
  {12}},\ \bibinfo {pages} {22451} (\bibinfo {year} {2022})}\BibitemShut
  {NoStop}%
\bibitem [{\citenamefont {Tersoff}(1988)}]{tersoff88}%
  \BibitemOpen
  \bibfield  {author} {\bibinfo {author} {\bibfnamefont {J.}~\bibnamefont
  {Tersoff}},\ }\bibfield  {title} {\bibinfo {title} {New empirical approach
  for the structure and energy of covalent systems},\ }\href
  {https://doi.org/10.1103/PhysRevB.37.6991} {\bibfield  {journal} {\bibinfo
  {journal} {Phys. Rev. B}\ }\textbf {\bibinfo {volume} {37}},\ \bibinfo
  {pages} {6991} (\bibinfo {year} {1988})}\BibitemShut {NoStop}%
\bibitem [{\citenamefont {Brenner}(1990)}]{brenner90}%
  \BibitemOpen
  \bibfield  {author} {\bibinfo {author} {\bibfnamefont {D.~W.}\ \bibnamefont
  {Brenner}},\ }\bibfield  {title} {\bibinfo {title} {Empirical potential for
  hydrocarbons for use in simulating the chemical vapor deposition of diamond
  films},\ }\href {https://doi.org/10.1103/PhysRevB.42.9458} {\bibfield
  {journal} {\bibinfo  {journal} {Phys. Rev. B}\ }\textbf {\bibinfo {volume}
  {42}},\ \bibinfo {pages} {9458} (\bibinfo {year} {1990})}\BibitemShut
  {NoStop}%
\bibitem [{\citenamefont {Pettifor}\ and\ \citenamefont
  {Oleinik}(1999)}]{pettifor99}%
  \BibitemOpen
  \bibfield  {author} {\bibinfo {author} {\bibfnamefont {D.~G.}\ \bibnamefont
  {Pettifor}}\ and\ \bibinfo {author} {\bibfnamefont {I.~I.}\ \bibnamefont
  {Oleinik}},\ }\bibfield  {title} {\bibinfo {title} {Analytic bond-order
  potentials beyond tersoff-brenner. i. theory},\ }\href
  {https://doi.org/10.1103/PhysRevB.59.8487} {\bibfield  {journal} {\bibinfo
  {journal} {Phys. Rev. B}\ }\textbf {\bibinfo {volume} {59}},\ \bibinfo
  {pages} {8487} (\bibinfo {year} {1999})}\BibitemShut {NoStop}%
\bibitem [{\citenamefont {Akagi}\ \emph {et~al.}(2012)\citenamefont {Akagi},
  \citenamefont {Udagawa},\ and\ \citenamefont {Motome}}]{akagi12}%
  \BibitemOpen
  \bibfield  {author} {\bibinfo {author} {\bibfnamefont {Y.}~\bibnamefont
  {Akagi}}, \bibinfo {author} {\bibfnamefont {M.}~\bibnamefont {Udagawa}},\
  and\ \bibinfo {author} {\bibfnamefont {Y.}~\bibnamefont {Motome}},\
  }\bibfield  {title} {\bibinfo {title} {Hidden multiple-spin interactions as
  an origin of spin scalar chiral order in frustrated kondo lattice models},\
  }\href {https://doi.org/10.1103/PhysRevLett.108.096401} {\bibfield  {journal}
  {\bibinfo  {journal} {Phys. Rev. Lett.}\ }\textbf {\bibinfo {volume} {108}},\
  \bibinfo {pages} {096401} (\bibinfo {year} {2012})}\BibitemShut {NoStop}%
\bibitem [{\citenamefont {Hayami}\ \emph {et~al.}(2017)\citenamefont {Hayami},
  \citenamefont {Ozawa},\ and\ \citenamefont {Motome}}]{hayami17}%
  \BibitemOpen
  \bibfield  {author} {\bibinfo {author} {\bibfnamefont {S.}~\bibnamefont
  {Hayami}}, \bibinfo {author} {\bibfnamefont {R.}~\bibnamefont {Ozawa}},\ and\
  \bibinfo {author} {\bibfnamefont {Y.}~\bibnamefont {Motome}},\ }\bibfield
  {title} {\bibinfo {title} {Effective bilinear-biquadratic model for
  noncoplanar ordering in itinerant magnets},\ }\href
  {https://doi.org/10.1103/PhysRevB.95.224424} {\bibfield  {journal} {\bibinfo
  {journal} {Phys. Rev. B}\ }\textbf {\bibinfo {volume} {95}},\ \bibinfo
  {pages} {224424} (\bibinfo {year} {2017})}\BibitemShut {NoStop}%
\bibitem [{\citenamefont {Furukawa}(1994)}]{furukawa94}%
  \BibitemOpen
  \bibfield  {author} {\bibinfo {author} {\bibfnamefont {N.}~\bibnamefont
  {Furukawa}},\ }\bibfield  {title} {\bibinfo {title} {Transport properties of
  the kondo lattice model in the limit $s=\infty$ and $d=\infty$},\ }\href
  {https://doi.org/10.1143/JPSJ.63.3214} {\bibfield  {journal} {\bibinfo
  {journal} {Journal of the Physical Society of Japan}\ }\textbf {\bibinfo
  {volume} {63}},\ \bibinfo {pages} {3214} (\bibinfo {year}
  {1994})}\BibitemShut {NoStop}%
\bibitem [{\citenamefont {Dagotto}\ \emph {et~al.}(1998)\citenamefont
  {Dagotto}, \citenamefont {Yunoki}, \citenamefont {Malvezzi}, \citenamefont
  {Moreo}, \citenamefont {Hu}, \citenamefont {Capponi}, \citenamefont
  {Poilblanc},\ and\ \citenamefont {Furukawa}}]{dagotto98}%
  \BibitemOpen
  \bibfield  {author} {\bibinfo {author} {\bibfnamefont {E.}~\bibnamefont
  {Dagotto}}, \bibinfo {author} {\bibfnamefont {S.}~\bibnamefont {Yunoki}},
  \bibinfo {author} {\bibfnamefont {A.~L.}\ \bibnamefont {Malvezzi}}, \bibinfo
  {author} {\bibfnamefont {A.}~\bibnamefont {Moreo}}, \bibinfo {author}
  {\bibfnamefont {J.}~\bibnamefont {Hu}}, \bibinfo {author} {\bibfnamefont
  {S.}~\bibnamefont {Capponi}}, \bibinfo {author} {\bibfnamefont
  {D.}~\bibnamefont {Poilblanc}},\ and\ \bibinfo {author} {\bibfnamefont
  {N.}~\bibnamefont {Furukawa}},\ }\bibfield  {title} {\bibinfo {title}
  {Ferromagnetic kondo model for manganites: Phase diagram, charge segregation,
  and influence of quantum localized spins},\ }\href
  {https://doi.org/10.1103/PhysRevB.58.6414} {\bibfield  {journal} {\bibinfo
  {journal} {Phys. Rev. B}\ }\textbf {\bibinfo {volume} {58}},\ \bibinfo
  {pages} {6414} (\bibinfo {year} {1998})}\BibitemShut {NoStop}%
\bibitem [{\citenamefont {Pekker}\ \emph {et~al.}(2005)\citenamefont {Pekker},
  \citenamefont {Mukhopadhyay}, \citenamefont {Trivedi},\ and\ \citenamefont
  {Goldbart}}]{pekker05}%
  \BibitemOpen
  \bibfield  {author} {\bibinfo {author} {\bibfnamefont {D.}~\bibnamefont
  {Pekker}}, \bibinfo {author} {\bibfnamefont {S.}~\bibnamefont
  {Mukhopadhyay}}, \bibinfo {author} {\bibfnamefont {N.}~\bibnamefont
  {Trivedi}},\ and\ \bibinfo {author} {\bibfnamefont {P.~M.}\ \bibnamefont
  {Goldbart}},\ }\bibfield  {title} {\bibinfo {title} {Double-exchange model
  for noninteracting electron spins coupled to a lattice of classical spins:
  Phase diagram at zero temperature},\ }\href
  {https://doi.org/10.1103/PhysRevB.72.075118} {\bibfield  {journal} {\bibinfo
  {journal} {Phys. Rev. B}\ }\textbf {\bibinfo {volume} {72}},\ \bibinfo
  {pages} {075118} (\bibinfo {year} {2005})}\BibitemShut {NoStop}%
\bibitem [{\citenamefont {Yunoki}\ \emph {et~al.}(1998)\citenamefont {Yunoki},
  \citenamefont {Hu}, \citenamefont {Malvezzi}, \citenamefont {Moreo},
  \citenamefont {Furukawa},\ and\ \citenamefont {Dagotto}}]{yunoki98}%
  \BibitemOpen
  \bibfield  {author} {\bibinfo {author} {\bibfnamefont {S.}~\bibnamefont
  {Yunoki}}, \bibinfo {author} {\bibfnamefont {J.}~\bibnamefont {Hu}}, \bibinfo
  {author} {\bibfnamefont {A.~L.}\ \bibnamefont {Malvezzi}}, \bibinfo {author}
  {\bibfnamefont {A.}~\bibnamefont {Moreo}}, \bibinfo {author} {\bibfnamefont
  {N.}~\bibnamefont {Furukawa}},\ and\ \bibinfo {author} {\bibfnamefont
  {E.}~\bibnamefont {Dagotto}},\ }\bibfield  {title} {\bibinfo {title} {Phase
  separation in electronic models for manganites},\ }\href
  {https://doi.org/10.1103/PhysRevLett.80.845} {\bibfield  {journal} {\bibinfo
  {journal} {Phys. Rev. Lett.}\ }\textbf {\bibinfo {volume} {80}},\ \bibinfo
  {pages} {845} (\bibinfo {year} {1998})}\BibitemShut {NoStop}%
\bibitem [{\citenamefont {Dagotto}(2003)}]{dagotto03}%
  \BibitemOpen
  \bibfield  {author} {\bibinfo {author} {\bibfnamefont {E.}~\bibnamefont
  {Dagotto}},\ }\href@noop {} {\emph {\bibinfo {title} {Nanoscale Phase
  Separation and Colossal Magnetoresistance}}}\ (\bibinfo  {publisher}
  {Springer Verlag},\ \bibinfo {year} {2003})\BibitemShut {NoStop}%
\bibitem [{\citenamefont {Ching}\ and\ \citenamefont {Huber}(1982)}]{ching82}%
  \BibitemOpen
  \bibfield  {author} {\bibinfo {author} {\bibfnamefont {W.~Y.}\ \bibnamefont
  {Ching}}\ and\ \bibinfo {author} {\bibfnamefont {D.~L.}\ \bibnamefont
  {Huber}},\ }\bibfield  {title} {\bibinfo {title} {Numerical studies of energy
  levels and eigenfunction localization in dilute three-dimensional systems
  with exponential interactions},\ }\href
  {https://doi.org/10.1103/PhysRevB.25.1096} {\bibfield  {journal} {\bibinfo
  {journal} {Phys. Rev. B}\ }\textbf {\bibinfo {volume} {25}},\ \bibinfo
  {pages} {1096} (\bibinfo {year} {1982})}\BibitemShut {NoStop}%
\bibitem [{\citenamefont {Logan}\ and\ \citenamefont {Winn}(1988)}]{logan88}%
  \BibitemOpen
  \bibfield  {author} {\bibinfo {author} {\bibfnamefont {D.~E.}\ \bibnamefont
  {Logan}}\ and\ \bibinfo {author} {\bibfnamefont {M.~D.}\ \bibnamefont
  {Winn}},\ }\bibfield  {title} {\bibinfo {title} {The density of states of a
  spatially disordered tight-binding model},\ }\href
  {https://doi.org/10.1088/0022-3719/21/34/013} {\bibfield  {journal} {\bibinfo
   {journal} {Journal of Physics C: Solid State Physics}\ }\textbf {\bibinfo
  {volume} {21}},\ \bibinfo {pages} {5773} (\bibinfo {year}
  {1988})}\BibitemShut {NoStop}%
\bibitem [{\citenamefont {Winn}\ and\ \citenamefont {Logan}(1989)}]{winn89}%
  \BibitemOpen
  \bibfield  {author} {\bibinfo {author} {\bibfnamefont {M.~D.}\ \bibnamefont
  {Winn}}\ and\ \bibinfo {author} {\bibfnamefont {D.~E.}\ \bibnamefont
  {Logan}},\ }\bibfield  {title} {\bibinfo {title} {A soluble theory for the
  density of states of a spatially disordered system},\ }\href
  {https://doi.org/10.1088/0953-8984/1/9/018} {\bibfield  {journal} {\bibinfo
  {journal} {Journal of Physics: Condensed Matter}\ }\textbf {\bibinfo {volume}
  {1}},\ \bibinfo {pages} {1753} (\bibinfo {year} {1989})}\BibitemShut
  {NoStop}%
\bibitem [{\citenamefont {Bush}\ \emph {et~al.}(1989)\citenamefont {Bush},
  \citenamefont {Logan}, \citenamefont {Madden},\ and\ \citenamefont
  {Winn}}]{bush89}%
  \BibitemOpen
  \bibfield  {author} {\bibinfo {author} {\bibfnamefont {I.~J.}\ \bibnamefont
  {Bush}}, \bibinfo {author} {\bibfnamefont {D.~E.}\ \bibnamefont {Logan}},
  \bibinfo {author} {\bibfnamefont {P.~A.}\ \bibnamefont {Madden}},\ and\
  \bibinfo {author} {\bibfnamefont {M.~D.}\ \bibnamefont {Winn}},\ }\bibfield
  {title} {\bibinfo {title} {The density of states of a spatially disordered
  system: theory compared with simulation},\ }\href
  {https://doi.org/10.1088/0953-8984/1/14/011} {\bibfield  {journal} {\bibinfo
  {journal} {Journal of Physics: Condensed Matter}\ }\textbf {\bibinfo {volume}
  {1}},\ \bibinfo {pages} {2551} (\bibinfo {year} {1989})}\BibitemShut
  {NoStop}%
\bibitem [{\citenamefont {Priour}(2012)}]{priour12}%
  \BibitemOpen
  \bibfield  {author} {\bibinfo {author} {\bibfnamefont {D.~J.}\ \bibnamefont
  {Priour}},\ }\bibfield  {title} {\bibinfo {title} {Electronic states in one-,
  two-, and three-dimensional highly amorphous materials: A tight-binding
  treatment},\ }\href {https://doi.org/10.1103/PhysRevB.85.014209} {\bibfield
  {journal} {\bibinfo  {journal} {Phys. Rev. B}\ }\textbf {\bibinfo {volume}
  {85}},\ \bibinfo {pages} {014209} (\bibinfo {year} {2012})}\BibitemShut
  {NoStop}%
\bibitem [{\citenamefont {Paszke}\ \emph {et~al.}(2019)\citenamefont {Paszke},
  \citenamefont {Gross}, \citenamefont {Massa}, \citenamefont {Lerer},
  \citenamefont {Bradbury}, \citenamefont {Chanan}, \citenamefont {Killeen},
  \citenamefont {Lin}, \citenamefont {Gimelshein}, \citenamefont {Antiga},
  \citenamefont {Desmaison}, \citenamefont {Kopf}, \citenamefont {Yang},
  \citenamefont {DeVito}, \citenamefont {Raison}, \citenamefont {Tejani},
  \citenamefont {Chilamkurthy}, \citenamefont {Steiner}, \citenamefont {Fang},
  \citenamefont {Bai},\ and\ \citenamefont {Chintala}}]{paszke19}%
  \BibitemOpen
  \bibfield  {author} {\bibinfo {author} {\bibfnamefont {A.}~\bibnamefont
  {Paszke}}, \bibinfo {author} {\bibfnamefont {S.}~\bibnamefont {Gross}},
  \bibinfo {author} {\bibfnamefont {F.}~\bibnamefont {Massa}}, \bibinfo
  {author} {\bibfnamefont {A.}~\bibnamefont {Lerer}}, \bibinfo {author}
  {\bibfnamefont {J.}~\bibnamefont {Bradbury}}, \bibinfo {author}
  {\bibfnamefont {G.}~\bibnamefont {Chanan}}, \bibinfo {author} {\bibfnamefont
  {T.}~\bibnamefont {Killeen}}, \bibinfo {author} {\bibfnamefont
  {Z.}~\bibnamefont {Lin}}, \bibinfo {author} {\bibfnamefont {N.}~\bibnamefont
  {Gimelshein}}, \bibinfo {author} {\bibfnamefont {L.}~\bibnamefont {Antiga}},
  \bibinfo {author} {\bibfnamefont {A.}~\bibnamefont {Desmaison}}, \bibinfo
  {author} {\bibfnamefont {A.}~\bibnamefont {Kopf}}, \bibinfo {author}
  {\bibfnamefont {E.}~\bibnamefont {Yang}}, \bibinfo {author} {\bibfnamefont
  {Z.}~\bibnamefont {DeVito}}, \bibinfo {author} {\bibfnamefont
  {M.}~\bibnamefont {Raison}}, \bibinfo {author} {\bibfnamefont
  {A.}~\bibnamefont {Tejani}}, \bibinfo {author} {\bibfnamefont
  {S.}~\bibnamefont {Chilamkurthy}}, \bibinfo {author} {\bibfnamefont
  {B.}~\bibnamefont {Steiner}}, \bibinfo {author} {\bibfnamefont
  {L.}~\bibnamefont {Fang}}, \bibinfo {author} {\bibfnamefont {J.}~\bibnamefont
  {Bai}},\ and\ \bibinfo {author} {\bibfnamefont {S.}~\bibnamefont
  {Chintala}},\ }\bibfield  {title} {\bibinfo {title} {Pytorch: An imperative
  style, high-performance deep learning library},\ }in\ \href
  {https://proceedings.neurips.cc/paper_files/paper/2019/file/bdbca288fee7f92f2bfa9f7012727740-Paper.pdf}
  {\emph {\bibinfo {booktitle} {Advances in Neural Information Processing
  Systems}}},\ Vol.~\bibinfo {volume} {32},\ \bibinfo {editor} {edited by\
  \bibinfo {editor} {\bibfnamefont {H.}~\bibnamefont {Wallach}}, \bibinfo
  {editor} {\bibfnamefont {H.}~\bibnamefont {Larochelle}}, \bibinfo {editor}
  {\bibfnamefont {A.}~\bibnamefont {Beygelzimer}}, \bibinfo {editor}
  {\bibfnamefont {F.}~\bibnamefont {d'Alch\'{e} Buc}}, \bibinfo {editor}
  {\bibfnamefont {E.}~\bibnamefont {Fox}},\ and\ \bibinfo {editor}
  {\bibfnamefont {R.}~\bibnamefont {Garnett}}}\ (\bibinfo  {publisher} {Curran
  Associates, Inc.},\ \bibinfo {year} {2019})\BibitemShut {NoStop}%
\bibitem [{\citenamefont {Barron}(2017)}]{barron17}%
  \BibitemOpen
  \bibfield  {author} {\bibinfo {author} {\bibfnamefont {J.~T.}\ \bibnamefont
  {Barron}},\ }\href@noop {} {\bibinfo {title} {Continuously differentiable
  exponential linear units}} (\bibinfo {year} {2017}),\ \Eprint
  {https://arxiv.org/abs/1704.07483} {arXiv:1704.07483 [cs.LG]} \BibitemShut
  {NoStop}%
\bibitem [{\citenamefont {Paszke}\ \emph {et~al.}(2017)\citenamefont {Paszke},
  \citenamefont {Gross}, \citenamefont {Chintala}, \citenamefont {Chanan},
  \citenamefont {Yang}, \citenamefont {DeVito}, \citenamefont {Lin},
  \citenamefont {Desmaison}, \citenamefont {Antiga},\ and\ \citenamefont
  {Lerer}}]{paszke17}%
  \BibitemOpen
  \bibfield  {author} {\bibinfo {author} {\bibfnamefont {A.}~\bibnamefont
  {Paszke}}, \bibinfo {author} {\bibfnamefont {S.}~\bibnamefont {Gross}},
  \bibinfo {author} {\bibfnamefont {S.}~\bibnamefont {Chintala}}, \bibinfo
  {author} {\bibfnamefont {G.}~\bibnamefont {Chanan}}, \bibinfo {author}
  {\bibfnamefont {E.}~\bibnamefont {Yang}}, \bibinfo {author} {\bibfnamefont
  {Z.}~\bibnamefont {DeVito}}, \bibinfo {author} {\bibfnamefont
  {Z.}~\bibnamefont {Lin}}, \bibinfo {author} {\bibfnamefont {A.}~\bibnamefont
  {Desmaison}}, \bibinfo {author} {\bibfnamefont {L.}~\bibnamefont {Antiga}},\
  and\ \bibinfo {author} {\bibfnamefont {A.}~\bibnamefont {Lerer}},\ }\bibfield
   {title} {\bibinfo {title} {Automatic differentiation in pytorch},\ }in\
  \href {https://openreview.net/forum?id=BJJsrmfCZ} {\emph {\bibinfo
  {booktitle} {NIPS 2017 Workshop on Autodiff}}}\ (\bibinfo {year}
  {2017})\BibitemShut {NoStop}%
\bibitem [{\citenamefont {Kingma}\ and\ \citenamefont {Ba}(2017)}]{kingma17}%
  \BibitemOpen
  \bibfield  {author} {\bibinfo {author} {\bibfnamefont {D.~P.}\ \bibnamefont
  {Kingma}}\ and\ \bibinfo {author} {\bibfnamefont {J.}~\bibnamefont {Ba}},\
  }\href@noop {} {\bibinfo {title} {Adam: A method for stochastic
  optimization}} (\bibinfo {year} {2017}),\ \Eprint
  {https://arxiv.org/abs/1412.6980} {arXiv:1412.6980 [cs.LG]} \BibitemShut
  {NoStop}%
\bibitem [{\citenamefont {Chern}\ \emph {et~al.}(2018)\citenamefont {Chern},
  \citenamefont {Barros}, \citenamefont {Wang}, \citenamefont {Suwa},\ and\
  \citenamefont {Batista}}]{chern18}%
  \BibitemOpen
  \bibfield  {author} {\bibinfo {author} {\bibfnamefont {G.-W.}\ \bibnamefont
  {Chern}}, \bibinfo {author} {\bibfnamefont {K.}~\bibnamefont {Barros}},
  \bibinfo {author} {\bibfnamefont {Z.}~\bibnamefont {Wang}}, \bibinfo {author}
  {\bibfnamefont {H.}~\bibnamefont {Suwa}},\ and\ \bibinfo {author}
  {\bibfnamefont {C.~D.}\ \bibnamefont {Batista}},\ }\bibfield  {title}
  {\bibinfo {title} {Semiclassical dynamics of spin density waves},\ }\href
  {https://doi.org/10.1103/PhysRevB.97.035120} {\bibfield  {journal} {\bibinfo
  {journal} {Phys. Rev. B}\ }\textbf {\bibinfo {volume} {97}},\ \bibinfo
  {pages} {035120} (\bibinfo {year} {2018})}\BibitemShut {NoStop}%
\bibitem [{\citenamefont {Bray}(1994)}]{bray94}%
  \BibitemOpen
  \bibfield  {author} {\bibinfo {author} {\bibfnamefont {A.~J.}\ \bibnamefont
  {Bray}},\ }\bibfield  {title} {\bibinfo {title} {Theory of phase-ordering
  kinetics},\ }\href {https://doi.org/10.1080/00018739400101505} {\bibfield
  {journal} {\bibinfo  {journal} {Advances in Physics}\ }\textbf {\bibinfo
  {volume} {43}},\ \bibinfo {pages} {357} (\bibinfo {year} {1994})}\BibitemShut
  {NoStop}%
\bibitem [{\citenamefont {Fisher}\ and\ \citenamefont {Huse}(1988)}]{fisher88}%
  \BibitemOpen
  \bibfield  {author} {\bibinfo {author} {\bibfnamefont {D.~S.}\ \bibnamefont
  {Fisher}}\ and\ \bibinfo {author} {\bibfnamefont {D.~A.}\ \bibnamefont
  {Huse}},\ }\bibfield  {title} {\bibinfo {title} {Nonequilibrium dynamics of
  spin glasses},\ }\href {https://doi.org/10.1103/PhysRevB.38.373} {\bibfield
  {journal} {\bibinfo  {journal} {Phys. Rev. B}\ }\textbf {\bibinfo {volume}
  {38}},\ \bibinfo {pages} {373} (\bibinfo {year} {1988})}\BibitemShut
  {NoStop}%
\bibitem [{\citenamefont {Huse}(1991)}]{huse91}%
  \BibitemOpen
  \bibfield  {author} {\bibinfo {author} {\bibfnamefont {D.~A.}\ \bibnamefont
  {Huse}},\ }\bibfield  {title} {\bibinfo {title} {Monte carlo simulation study
  of domain growth in an ising spin glass},\ }\href
  {https://doi.org/10.1103/PhysRevB.43.8673} {\bibfield  {journal} {\bibinfo
  {journal} {Phys. Rev. B}\ }\textbf {\bibinfo {volume} {43}},\ \bibinfo
  {pages} {8673} (\bibinfo {year} {1991})}\BibitemShut {NoStop}%
\end{thebibliography}%

\end{document}